\documentclass[letterpaper,onecolumn,11pt,accepted=2025-03-02]{quantumarticle}
\pdfoutput=1
\usepackage[utf8]{inputenc}
\usepackage[english]{babel}
\usepackage{amsmath}
\usepackage{amssymb}
\usepackage{amsthm}
\usepackage{stmaryrd}
\usepackage{hyperref}
\usepackage{caption,subcaption}
\usepackage[numbers,sort&compress]{natbib}
\usepackage{tikz}
\usepackage{lipsum}
\usepackage{braket}
\usepackage{ulem}

\newcommand{\h}{{\mathcal H}}

\renewcommand{\r}{{\rm R}}

\newcommand{\s}{{\rm S}}
\newcommand{\z}{{\rm Z}}

\newcommand{\bbbone}{\mathchoice {\rm 1\mskip-4mu l} {\rm 1\mskip-4mu l}
{\rm 1\mskip-4.5mu l} {\rm 1\mskip-5mu l}}
\newtheorem{thm}{Theorem}

\newtheorem{lem}{Lemma}

\begin{document}

\title{Ultrastrong coupling, nonselective measurement and quantum Zeno dynamics}

\author{Stefano Marcantoni}
\email{stefano.marcantoni@univ-cotedazur.fr}
\affiliation{Laboratoire J. A. Dieudonn\'e (LJAD), Universit\'e C\^ote d'Azur, Parc Valrose, 06108 Nice, France}

\author{Marco Merkli}
\email{merkli@mun.ca}
\affiliation{Department of Mathematics and Statistics, Memorial University of Newfoundland, St.~John's, NL, A1C 5S7 Canada}

\maketitle

\begin{abstract} 
We study the dynamics of an open quantum system linearly coupled to a bosonic reservoir. We show that, in the ultrastrong coupling limit, the system undergoes a nonselective measurement and then evolves unitarily according to an effective Zeno Hamiltonian. This dynamical process is largely independent of the reservoir state. We examine the entanglement breaking effect of the ultrastrong coupling on the system. We also derive the evolution equation for systems in contact with several reservoirs when one coupling is ultrastrong. The effective system dynamics displays a rich structure and, contrarily to the single reservoir case, it is generally non-Markovian. Our approach is based on a Dyson series expansion, in which we can take the ultrastrong limit termwise, and a subsequent resummation of the series. Our derivation is mathematically rigorous and uncomplicated.
\end{abstract}

\section{Introduction}

An open quantum system is modeled as a bipartite system-reservoir ($\s\r$) complex, described by a Hilbert space
$$
\h =\h_\s\otimes\h_\r ,
$$
and has a Hamiltonian of the form 
\begin{equation*}
H = H_\s + H_\r +\lambda G\otimes\varphi.
\end{equation*}
Here, $H_\s$ and $H_\r$ are the system and reservoir Hamiltonians, and $G$, $\varphi$ are hermitian operators on $\h_\s$ and $\h_\r$ determining the interaction. The parameter $\lambda\in\mathbb R$ is a coupling constant. It is understood that $\s$ is a small system, which for us here means that $\dim\h_\s<\infty$, and that $\r$ is a large system, which means here that $\dim\h_\r=\infty$ and that $H_\r$ has a continuum of modes (continuous spectrum). The reduced system state is given by the density matrix
$$
\rho_\s(t) = {\rm tr}_\r \,\rho_\s\otimes\omega_\r \big(e^{it H}\cdot e^{-itH}\big),
$$
where $\rho_\s\otimes\omega_\r$ is the initial (factorized) system-reservoir state and ${\rm tr}_\r$ denotes the partial trace over the reservoir degrees of freedom. As we explain below, $\omega_\r$ can be a density matrix of $\r$ | but should generally be understood as an `expectation functional'.

\begin{figure}[h!]
\centering
\includegraphics[width=.96\textwidth]{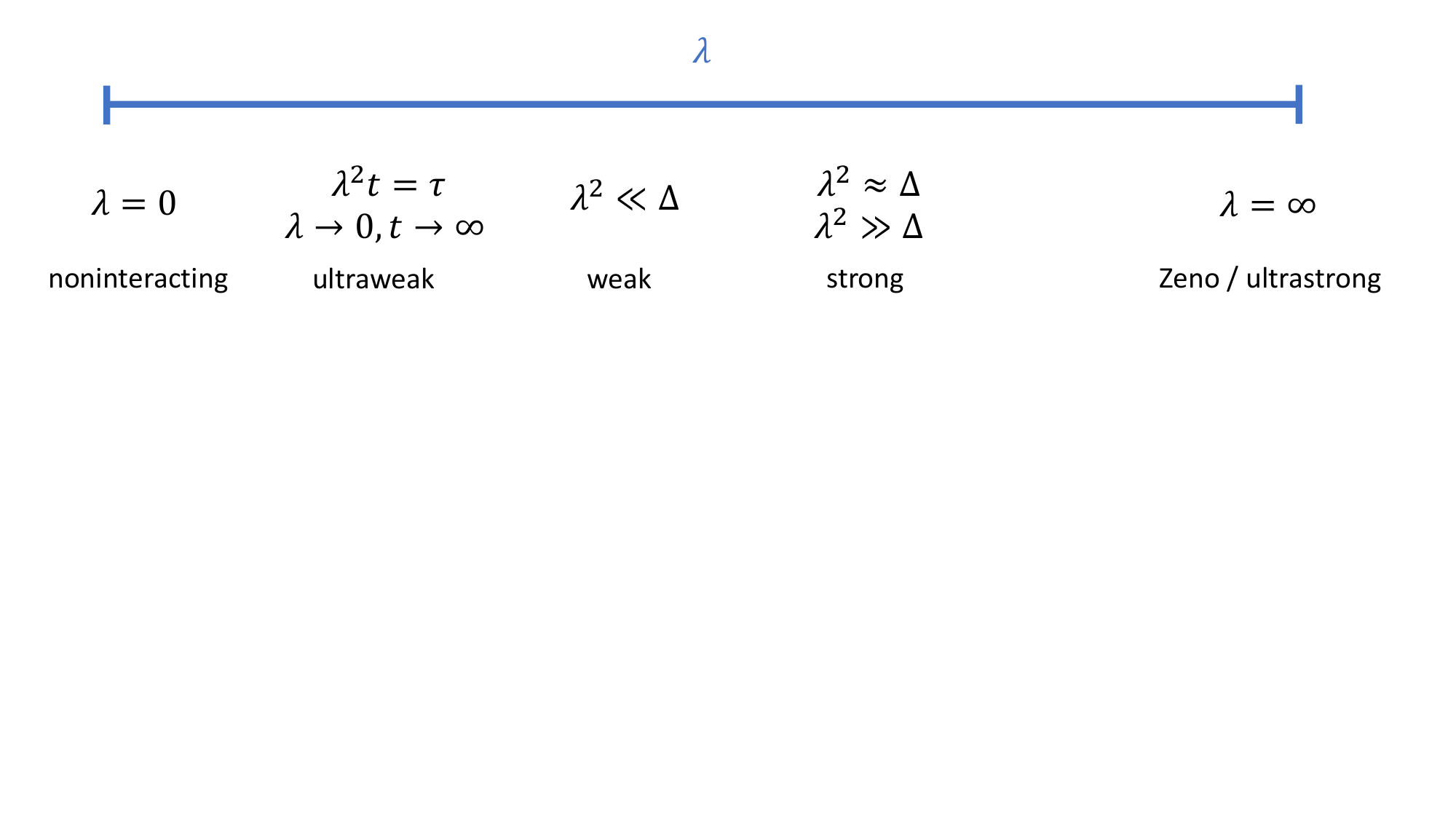}
\vspace*{-5.0cm}
\caption{Different coupling regimes: From noninteracting to ultrastrong. Here, $\Delta$ represents a typical system Bohr frequency. Energy shifts in $\s$ due to the interaction with $\r$ are typically $\propto \lambda^2$. In all regimes one can consider static and dynamical properties of the system, except for the ultraweak coupling regime. That one is a purely dynamical regime in which time is coarse grained to the value $\tau=\lambda^2t$. In the current paper we focus on the ultrastrong coupling regime $\lambda\rightarrow\infty$. We show that it gives rise to a nonselective measurement of the system followed by a quantum Zeno dynamics.}
\label{Fig:Regime} 
\end{figure}

Of course, $\rho_\s(t)$ depends on the coupling parameter $\lambda$ and a key question is how the properties of the system depend on the strength of $\lambda$. The two extreme cases $\lambda=0$ and `$\lambda=\infty$' are called the {\it noninteracting} and the {\it ultrastrong coupling regimes}, respectively. In the first case, the system $\s$ does not interact with the reservoir $\r$ and it evolves separately, undisturbed by the environment. The case of infinitely strong (ultrastong) coupling is understood in a limit sense when $|\lambda|\rightarrow\infty$. Between the two extreme cases one finds other regimes. The {\it ultraweak} coupling regime, also called the {\it van Hove weak coupling regime}, is defined by taking the limit $\lambda\rightarrow 0$ and at the same time, looking at the dynamics for long times $t$, such that $\lambda^2t=\tau$ takes finite values \cite{VH55,FP99}. The intuition is that for weaker interaction strength, one has to wait for a longer time to detect a sizable influence on the system caused by the reservoir | noticeable interaction effects happen over a {\it coarse grained} time scale parametrized by $\tau$. One can show that the Markovian approximation for the reduced system dynamics is valid for certain systems in the ultraweak coupling regime | leading to an effective approximate evolution given by the Markovian master equation. This was first proved in  \cite{Da74,Da76} with subsequent refinements  \cite{Rivas17,Rivas10}. If $\lambda$ is small but fixed (without taking $\lambda\rightarrow 0$) then we are said to be in the {\it weak coupling regime}. The smallness of $\lambda$ is taken in comparison for example with the Bohr energies of $\s$ (which are the nonvanishing energy eigenvalue differences of $H_\s$). In this regime, too, the correctness of the Markovian master equation can be validated for certain models \cite{MMAOP,MMQI,MMQII,MM22}, using the so called quantum resonance theory. The Markovian regime is particularly suited to describe, for example, matter-radiation interactions and quantum optical systems.  

Towards the other extreme, the {\it strong coupling regime} is characterized by values of $\lambda$ which are large compared to the `hopping' terms in $H_\s$, namely the part of $H_\s$ which does not commute with the interaction operator $G$. In this regime, one can perform a `polaron transformation' after which the $\s\r$ dynamics is explicitly solvable in the absence of the hopping terms (but in which $\s$ and $\r$ are coupled with arbitrary strength $\lambda$). The hopping terms are then treated as a perturbation. This strategy is particularly adapted to the treatment of, for example, the F\"orster and Marcus theories, describing the excitation energy and charge transfers in quantum chemical and biological processes \cite{Foe48, Marcus56, MK11,Tru22,M14,M16}. A further approach to describe the strong coupling regime is the reaction coordinate method, in which one incorporates some degrees of freedom of the reservoir into the system \cite{Stras16,Segal23a, Segal23b}. The (ultra)strong coupling regime is of interest for static, not only  dynamical, properties as well in particular in quantum equilibrium and non-equilibrium thermodynamics \cite{TMCA22,CA21,Rivas20,GK19,MA17}.

While there are somewhat general mathematical approaches to describe the ultraweak coupling regime (and the non-interacting one), the mentioned techniques to treat the weak and strong coupling settings are tailored to specific models and a general solution of the problem is not available. A discussion of this fact can be found for instance in \cite{TMCA22}. In the following, we focus on the ultrastrong coupling limit and present a detailed analysis of the open system dynamics in this regime.

\medskip

There is another, {\it a priori} quite different perspective on the theory of strong interactions, coming from the study of the quantum Zeno effect. This effect describes the fate of a system subjected to frequent measurements. It is formalized as follows. A system's density matrix $\rho_\s$ evolves unitarily, according to a Hamiltonian $H_\s$,
$$
\rho_\s\mapsto \mathcal U_t(\rho_\s) =e^{-it H_\s} \rho_\s\, e^{it H_\s}.
$$
Let $\{P_n\}$ be a complete set of orthogonal projections ($P_nP_m=\delta_{mn}P_n$, $\sum_n P_n=1$) | think of the $\{P_n\}$ as the spectral projections of a system observable to be measured. Those projections describe the non-selective measurement on a density matrix of $\s$ by
$$
\rho_\s \mapsto {\mathcal P}(\rho_\s)=\sum_n P_n\rho_\s P_n.
$$
The evolution of $\s$ interspersed with $N$ measurements at time intervals of duration $t/N$ each, is 
$$
\rho_\s(t,N) = \big(\mathcal P\, \mathcal U_{t/N} \big)^N\rho_\s.
$$
It is then shown that \cite{MS76,FP08,Fetal2000,EI21}
$$
\lim_{N\rightarrow\infty}\rho_\s(t,N) = e^{-it H_\z}\big( \sum_n P_n\rho_\s P_n\big)  e^{it H_\z},
$$
where the right hand side is called the {\it Zeno dynamics}, with the {\it Zeno Hamiltonian} given by
$$
H_\z = \sum_n P_n H_\s P_n.
$$
The action of the frequent measurement is thus to cut in $\rho_\s$ correlations between different subspaces ${\rm Ran}P_n$ | called the {\it Zeno subspaces} | and to evolve each one independently by the projected Hamiltonian $P_n H_\s P_n$. While the dynamics within a block is still carrying a mark of the Hamiltonian $H_\s$, the partitioning of the space into blocks is determined entirely by the measurement projections $P_n$. 
\medskip

The basic intuition for the {\bf connection between the Zeno and the ultrastrong coupling} is that the quantum von  Neumann measurements are supposed to happen instantaneously, and the associated very short (zero) time would correspond to a very strong (infinitely strong) interaction with an apparatus. In the above description of the Zeno effect of a system $\s$, however, there is no explicit mention of an apparatus, or reservoir $\r$. The question then is,
\begin{itemize}
    \item[{ Q:}] {\bf Does an ultrastrong $\s\r$ coupling cause a nonselective measurement and subsequent Zeno dynamics?}
\end{itemize}
Our contribution in the current work is to  answer this question in the positive for a large class of open systems where a finite dimensional $\s$ is coupled to a reservoir of bosonic modes $k\in\mathbb R^3$ with creation and annihilation operators satisfying $[a(k),a^\dag(l)]=\delta(k-l)$. We consider  interaction operators of the form $G\otimes\varphi(g)$, where $G$ is a hermitian observable of $\s$ and $\varphi(g)$ is the field operator,
$$
G=\sum_n \gamma_n P_n,\qquad \varphi(g) = \frac{1}{\sqrt 2} \int_{\mathbb R^3} \big( g(k) a^\dag(k) + {\rm h.c.} \big)d^3k,
$$
the $\gamma_n$ (which are distinct for distinct $n$) and $P_n$ are the eigenvalues and eigenprojections of $G$, and  $g\in L^2({\mathbb R}^3,d^3k)$ is called the form factor, determining how strongly each mode $k$ is coupled to $\s$. This class of models includes the famous spin-Boson model. Let $\rho_\s(t)={\rm Tr}_\r \,\rho_\s\otimes\omega_\r (e^{it H}\cdot e^{-itH})$ be the reduced system density matrix, obtained by tracing out the reservoir, as introduced above. Our results are as follows.\vspace{0.2cm}\\
$\bullet$ Our {\bf main result} is that for all times $t>0$,
$$
\lim_{|\lambda|\rightarrow\infty} \rho_\s(t) = e^{-it H_\z}\Big( \sum_n P_n\rho_\s P_n\Big)  e^{it H_\z}
$$
with Zeno Hamiltonian $H_\z=\sum_n P_n H_\s P_n$. Our result holds for initial states $\omega_\r$ of the reservoir drawn from a large class | including all Gaussian states, such as equilibrium states at any temperature. The ultrastrong coupling limit has an effect on the system (right hand side of the above equation) which is independent of the reservoir state. This answers the above question:
\begin{itemize}
\item[{ A:}] {\bf The ultrastrong coupling implements a nonselective measurement of the system coupling operator $G$ and a subsequent associated Zeno dynamics.}
\end{itemize}
We complement this main finding with further results:
\smallskip

\noindent
$\bullet$ When $\s$ is a many-body system we show that for typical choices of the coupling operator $G$, the ultrastrong coupling limit breaks entanglement between the subunits of the system.
\smallskip

\noindent
$\bullet$ We find the dynamics of complexes when $\s$ is coupled to several reservoirs,  when one of them is coupled ultrastrongly to $\s$. We show that this ultrastrong coupling causes a Zeno dynamics for $\s$ and the residual reservoirs. The reduction to $\s$ gives a rich, generally non Markovian dynamics | in contrast to the case of a single reservoir. 
\medskip

{\it Related previous work on the Zeno effect in open systems.} In the previous literature, one line of investigation examines the frequent measurements setup  with the unitary dynamics $\mathcal U_t$ replaced by the action of a CPTP (completely positive, trace preserving) semigroup generated by operators in the standard GKSL (Gorini-Kossakovski-Sudarshan-Lindblad) form \cite{Bur19}. That approach differs from ours because it considers the effect of the reservoir implicitly | the reservoir is already `traced out' before the frequent measurements are performed | and it is assumed that the system dynamics is {\it Markovian} (GKSL semigroup). This is a different physical model from ours as we investigate a microscopic $\s\r$ model and show that the ultrastrong coupling produces a Zeno effect on $\s$. There is a remarkable current activity in the mathematical analysis aimed at finding explicit error bounds for the deviation of the Zeno dynamics from the dynamics caused by frequent (but not infinitely frequent) measurements, under generic assumptions on the structure of the semigroups. See for instance the works \cite{Bur19,Bur22,MR23,Sal24}.

Some literature is more in line with our approach, where the reservoirs are explicitly included in the description. In \cite{Tru22} the author derives a Markovian master equation in the strong coupling limit, starting from a system-bath model. They exhibit the emergence of the Zeno dynamics and corrections to it (large but finite $\lambda$). Their method is based on second order perturbation theory in the hopping terms in $H_\s$, with uncontrolled higher order terms. 
In \cite{PS97} the authors analyze an unstable (3-level) system interacting with a radiation field (oscillators) and they compare the decay, or de-excitation rates (Fermi Golden Rule) of the system in the presence and in the absence of an additional strong laser field illuminating the system (the measurement apparatus), revealing that the decay is slowed down by the laser, in accordance with the Zeno effect. In \cite{FP2000} the authors find that the decay rates of a three level system can also be enhanced, depending on which level transitions the laser couples strongly to (`inverse' or anti-Zeno effect). In \cite{Frasca} a Schr\"odinger particle in a Coulomb field subjected to a strong interaction with a monochromatic electromagnetic wave is analyzed. The author shows that for small times, the decay of the particle follows a law consistent with the Zeno effect (as opposed to an exponential decay). In \cite{Dvira07} the authors consider the spin-boson model (discrete modes) and analyze the short-time decay rates of the initially excited spin. They detect that depending on the strength of the interaction with the bath, the decay rate is decreasing or increasing in the coupling parameter | hence revealing a Zeno or anti-Zeno behaviour. In \cite{POT} quantum measurement processes were studied in view of the Zeno effect, and connected to experiments on inhibited stimulated emissions in a matter-radiation systems. In comparison with these works, our results are quite general (valid and independent for a large class of reservoir states) and are mathematically rigorous and quite uncomplicated. That said, so far we only consider $|\lambda|\rightarrow \infty$, which is a simpler regime than $\lambda$ finite but large. Work on that harder regime is under way.

\section{Setup and main results}
\subsection{The model}

A $d$-level quantum system is coupled to a reservoir with a continuum of modes. Each mode is labeled by $k\in\mathbb R^3$ and has an associated bosonic creation and annihilation operators $a^\dag(k),a(k)$, satisfying the canonical commutation relation $[a(k),a^\dag(l)]=\delta(k-l)$. The total Hamiltonian is given by
\begin{equation}
\label{Ham}
H=H_\s +H_\r +\lambda G\otimes \varphi(g),
\end{equation}
where $H_\s$ is the system Hamiltonian, that is a $d\times d$ hermitian matrix and 
\begin{equation}
\label{HR}
H_\r = \int_{\mathbb R^3} \omega(k) a^\dag(k)a(k) d^3k
\end{equation}
is the reservoir Hamiltonian. In \eqref{HR}, $\omega(k)\ge 0$ is the energy of the mode $k$ (`dispersion relation'). The interaction term in \eqref{Ham} carries a {\it coupling constant} $\lambda\in\mathbb R$, a system coupling operator 
\begin{equation}
\label{G}
G = \sum_{l=1}^\nu \gamma_l P_l
\end{equation}
with distinct (possibly degenerate) eigenvalues $\gamma_l$ and spectral projections $P_l$ ($\dim P_l\ge 1$), and the field operator 
\begin{equation}
\label{field}
\varphi(g) = \frac{1}{\sqrt 2} \int_{\mathbb R^3} \big( g(k) a^\dag(k) + {\rm h.c.} \big)d^3k,
\end{equation}
where $g(k)\in L^2({\mathbb R}^3,d^3k)$ is a complex valued function, called the form factor.

In the physics literature reservoirs are often taken to be collections of harmonic oscillators with {\it discrete} energy spectrum. This amounts to replacing the integral in \eqref{HR} by a sum over mode energies $\omega_k$, with say $k\in\mathbb N$. However, in order to describe physical phenomena such as decoherence, thermalization or generally irreversible dynamics, one needs to take a continuous mode limit. In the current work, we start off directly with a continuum mode reservoir.

The Hamiltonian $H_\r$ of the reservoir, \eqref{HR}, has purely absolutely continuous spectrum starting at $\inf_k \omega(k)$ and as a consequence, $e^{-\beta H_\r}$ even though well defined as a bounded operator, is not trace class. Namely\footnote{As is well known from basic theory of operators, if $e^{-X}$ is trace class for an operator $X=X^\dag$, then $e^{-X}$ must be a compact operator, which in turn means that $e^{-X}$ must have discrete eigenvalues which can accumulate at the point zero only. This implies that $X$ must have purely discrete eigenvalues which may grow to $\infty$, but $X$ cannot have continuous spectrum.
}, ${\rm tr}_\r\, e^{-\beta H_\r}=\infty$. This implies that one must give an alternative expression for the equilibrium state of the reservoir, other than the `Gibbs density matrix' $\propto e^{-\beta H_\r}$. The construction of the continuous mode equilibrium state is done by taking a limit of discrete mode equilibrium states (the `thermodynamic limit', see e.g. \cite{BR,MM06}). It results in an expectation functional $\omega_{\r,\beta}$ for reservoir observables (built as functions of $a^\dag(k)$, $a(l)$), which can be expressed entirely by the {\it characteristic function}
\begin{equation}
\label{charfun}
\omega_{\r,\beta}(W(f)) =e^{-\frac14 \langle f,\coth(\beta\omega/2)f\rangle}.
\end{equation}
Here, $\langle f,h\rangle = \int_{\mathbb R^3} \overline{f(k)} h(k) d^3k$ is the inner product of $L^2({\mathbb R^3}, d^3k)$ and $W(f)$ is the unitary {\it Weyl operator},
\begin{equation*}
W(f) =e^{i\varphi(f)}
\end{equation*}
with $\varphi(f)$ as in \eqref{field}. The characteristic function \eqref{charfun} is also called the generating function, as it can be used to express the expectation for any observable by using the relation $\varphi(f) = -i\partial_\alpha|_{\alpha =0} W(\alpha f)$. One then finds that the two-point function of the reservoir equilibrium state is given by
\begin{equation}
\label{tpfun}
\omega_{\r,\beta}\big(a^\dag(k)a(l)\big) = \frac{\delta(k-l)}{e^{\beta\omega(k)}-1}.
\end{equation}
This encodes Planck's law of black body radiation, where $n(k) = (e^{\beta\omega(k)}-1)^{-1}$ is the momentum density distribution in the reservoir | that is, the number of modes per unit volume (in space $x\in\mathbb R^3)$ in a given momentum region $\Lambda\in\mathbb R^3$ is $\int_{\Lambda} n(k) d^3k$. The state \eqref{charfun} is Gaussian and centered. Its covariance operator $\mathcal C$ acting on $L^2(\mathbb R^3,d^3k)$ is the multiplication by the function $\coth(\beta\omega/2)$ operator. We consider more general Gaussian states of the form
\begin{equation}
\label{gaus}
\omega_\r(W(f)) = e^{-\frac14 \langle f , \mathcal C f\rangle},
\end{equation}
where $\mathcal C$ is an operator on $L^2(\mathbb R^3,d^3k)$, satisfying
\begin{equation}
\label{C>1}
\mathcal C\ge \bbbone.
\end{equation}
The condition \eqref{C>1} is known to be necessary and sufficient for the right hand side of \eqref{gaus} to be the expectation functional of a quantum state | the case $\mathcal C=\bbbone$ is the field vacuum ($T=0$ temperature) case.  Instead of the thermal distribution \eqref{tpfun} we may consider reservoir states with an arbitrary energy distribution $\mu(k)\ge 0$, $\omega_\r\big(a^\dag(k)a(l)\big) = \mu(k) \delta(k-l)$, 
which corresponds to the covariance operator being the multiplication by the function $C(k) = 1+2\mu(k)$, compare with \eqref{charfun}. The corresponding state $\omega_\r$ (see \eqref{gaus}) is stationary: $\omega_\r(e^{it H_\r}W(f)e^{-it H_\r}) = \omega_\r(W(e^{it\omega}f))=\omega_\r(W(f))$. Covariance matrices $\mathcal C$ which are not multiplication operators by a function of $k$ result in non-stationary  Gaussian reservoir states, and they are included in our theory. Our result holds as well for non-centered Gaussian states; an example is the coherent state 
$\omega_{\r,\alpha}(W(f)) = \langle W(\alpha)\Omega |  W(f)W(\alpha)\Omega\rangle$, where $\Omega$ is the vaccum vector and $\alpha\in L^2(\mathbb R^3,d^3k)$ is fixed, and whose characteristic functional is $\omega_{\r,\alpha}(W(f)) = e^{-\frac14\langle f,f\rangle+i {\rm Im}\langle \alpha,f\rangle}$. Another example where our results apply is a reservoir in equilibrium including a condensate, for which the expectation functional is given by the product of a centered Gaussian with a Bessel function \cite{MM06} (we can treat the case of a Gaussian multiplied by any bounded function of $f$). For ease of presentation, we simply assume \eqref{gaus}. 
\medskip

We take initial system-reservoir states of the form\footnote{ The product initial state is typically assumed when describing open quantum system dynamics, sometimes even in the strong coupling setting (see e.g. \cite{PS97, FP2000}). One might argue that in the latter case this assumption is not always realistic and we plan to remove it in a following work. } 
\begin{equation*}
\omega_{\s\r} = \omega_\s\otimes\omega_\r,
\end{equation*}  
where $\omega_\r$ is the Gaussian state \eqref{gaus} for a general covariance operator $\mathcal C\ge \bbbone$, and where 
\begin{equation*}
\omega_\s (\cdot) = {\rm tr}_\s \big(\rho_\s \cdot \big)
\end{equation*}
is a system state determined by a density matrix $\rho_\s$ of the $d$-level system with Hilbert space $\mathbb C^d$. Let $A\in\mathcal B(\mathbb C^d)$ (bounded operators) be a system observable. The reduced system density matrix $\rho_\s(t)$ at time $t\ge 0$ in the ultrastrong coupling limit is defined by the relation
\begin{equation}
\label{rhos}
{\rm tr}_\s\big( \rho_\s(t)  A\big) = \lim_{|\lambda|\rightarrow\infty} \omega_\s\otimes\omega_\r
\big(e^{i t H} (A\otimes\bbbone_\r) e^{-it H}\big),
\end{equation}
holding for all system observables $A\in\mathcal B(\mathbb C^d)$.
\medskip

{\bf Assumptions on $g$ and $\omega$.} The form factor $g=g(k)\in L^2(\mathbb R^3,d^3k)$ and the dispersion relation $\omega(k)\ge 0$ are quite general here. We assume that for all $t\ge 0$,
\begin{equation}
\label{cond1}
\mathcal C^{1/2} \,\frac{e^{i\omega(k)t}-1}{\omega(k)}g(k)\in L^2(\mathbb R^3,d^3k).
\end{equation}
This is a joint condition on $g,\omega$ and the covariance $\mathcal C$ of the state $\omega_\r$, \eqref{gaus}. As $\mathcal C\ge \bbbone$ its square root $\mathcal C^{1/2}$ is a well defined (positive) operator. Typically, $\mathcal C$ is not a bounded operator (this is so even in the thermal case, see the discussion above) and \eqref{cond1} means that the function $\frac{e^{i\omega t}-1}{\omega}g$ must belong to the domain of $\mathcal C^{1/2}$. We further require a bit of regularity, namely that there is a $k_0\in\mathbb R^3$ such that $g(k_0)\neq 0$ and $g$ is continuous in a neighbourhood $\mathcal N_0$ of $k_0$ (open ball centered at $k_0$). Furthermore, we assume that the set of values of $\omega(k)$,  as $k$ varies throughout $\mathcal N_0$, either contains the value zero ($\exists k_0\in\mathcal N_0$ such that $\omega(k_0)=0$) or it contains two points $\omega_1,\omega_2$ such that $\omega_2/\omega_1\not\in\mathbb Q$. In particular, it suffices that the values of $\omega(k)$ contain any interval $(a,b)\subset\mathbb R$ as $k$ varies throughout $\mathcal N_0$. This condition is very little restrictive, even though it may appear unusual | we use it to make sure that certain terms multiplying $|\lambda|$ do not vanish and hence drive the decay in the ultrastrong coupling limit (see Lemma \ref{lem:2} below). 

The condition \eqref{cond1} may require properties of $g$ other than $g\in L^2(\mathbb R^3,d^3k)$. This happens in particular if the covariance is unbounded (as {\it e.g.} in the thermal case, \eqref{charfun} for $\omega\rightarrow 0$). If $\omega(k)$ is a radial function, that is, depending on $|k|$ only, then $\omega(|k|)$ may have discontinuities, gaps, jumps or cusps in its graph, and it may vanish at points or in intervals of $|k|$ | these cases are covered by our assumptions. And so are of course the typical examples $\omega(k)=|k|, |k|^2, \sqrt{|k|^2+1}$.
\medskip

{\bf Notation.} In the sequel, we will often write $\omega$ instead of $\omega(k)$ for the dispersion relation. We caution that the similar symbols $\omega_\s$ and $\omega_\r$ denote states of the system and the reservoir.

\subsection{Ultrastrong coupling gives Zeno dynamics}

The following is our first main result.

\begin{thm}
\label{thm1}
For all $t > 0$, the system density matrix \eqref{rhos} in the ultrastrong coupling limit is given by
\begin{equation}
\label{dyns}
\rho_\s(t) = e^{-it H_\z} \left( \sum_{l=1}^\nu P_l  \rho_\s  P_l \right) e^{it H_\z},
\end{equation}
where $H_\z$ is the Zeno Hamiltonian (see \eqref{G})
\begin{equation}
\label{HZeno}
H_\z = \sum_{l=1}^\nu P_l H_\s P_l.
\end{equation}
\end{thm}
Theorem \ref{thm1} is a direct consequence of the following more general result for the dynamics of both $\s$ {\bf and} $\r$.
\begin{thm}
\label{thm2}
Let $A\in\mathcal B(\h_\s)$ and $h\in L^2({\mathbb R}^3,d^3k)$ such that $\mathcal C^{1/2}h \in L^2({\mathbb R}^3,d^3k)$, where we recall that $\mathcal C$ is the covariance \eqref{gaus}. We have for all $t>0$, 
\begin{align}
\lim_{|\lambda|\rightarrow\infty}\omega_\s\otimes\omega_\r &\Big( e^{itH} \big\{\big(e^{\frac i2\lambda G {\rm Re}\langle g, \frac{1-e^{i\omega t}}{\omega} h\rangle} Ae^{\frac i2\lambda G {\rm Re}\langle g, \frac{1-e^{i\omega t}}{\omega} h\rangle}\big)\otimes W(h)\big\}e^{-itH}\Big)\nonumber\\
&= \sum_{l=1}^\nu \omega_\s\big(e^{it H_{\rm Z}} P_l A P_l e^{-it H_{\rm Z}}\big) \omega_\r\big(W(e^{i\omega t}h)\big),
\label{thmeqn}
\end{align}
where the Zeno Hamiltonian $H_\z$ is given in \eqref{HZeno}. 
\end{thm}
We give a proof of Theorem \ref{thm2} in Section \ref{sec:proofthm2}. The equation  \eqref{thmeqn} shows that the full $\s\r$ dynamics has eternal (very fast) oscillations in the limit $|\lambda|\rightarrow\infty$. Only the reduction to $\s$ alone ($h=0)$ yields a true limit \eqref{rhos}, \eqref{dyns}. The system dynamics \eqref{dyns} does not depend at all on the form factor $g(k)$, the dispersion relation $\omega(k)$ or the reservoir state encoded by the covariance $\mathcal C$. It is entirely determined by the system Hamiltonian $H_\s$ and the interaction operator $G$. The `regularity condition' \eqref{cond1} is not needed for $\rho_\s(t)$, \eqref{dyns} to make sense | it is only used in its derivation.  In the literature on the renormalizability of the spin Boson model and its generalizations one tries to make sense of Hamiltonians having singular form factors $g$, which do not belong to $L^2$, for example because they are infrared or ultraviolet singular \cite{DM20,LN22,LN23}. It would be interesting to investigate if those singular Hamiltonians lead to the same ultrastrong coupling result we derive.

\subsection{Ultrastrong coupling breaks entanglement within $\s$} 

Consider a system $\s$ consisting of $N$ subsystems $\s_1,\ldots,\s_N$,  described by the Hilbert space 
\begin{equation*}
\h_\s=\h_{\s, 1}\otimes\cdots \otimes\h_{\s, N},
\end{equation*}
possibly of varying finite subsystem dimensions. To each subsystem $n$, we associate a hermitian interaction operator $G_n$ acting on $\h_{\s,n}$ and we write for simplicity, 
$$
G_n\equiv \bbbone\otimes\cdots\otimes G_n\otimes\cdots\otimes \bbbone.
$$
Wen consider total interaction operators of the form 
\begin{equation}
\label{HSR}
H_{\rm SR} = G\otimes\varphi(g),\qquad G = F(G_1,\ldots,G_N), 
\end{equation}
where $F:{\mathbb R}^N\rightarrow\mathbb R$ is a continuous function of $N$ variables. Examples are,
$$
F(G_1,\ldots,G_n) = \sum_{n=1}^N G_n,\qquad \mbox{or}\qquad F(G_1,\ldots,G_N) =G_1\cdots G_N.
$$
(However, see \eqref{notG} for reasonable interactions which are not of this form | and which lead to outcomes different from the ones discussed here.) 
The $\s\r$ Hamiltonian is given by
\begin{equation}
\label{totHamilt}
H = H_\s +H_\r +\lambda H_{\s\r},
\end{equation}
where $H_\s$ is a hermitian operator on $\h_\s$, $H_\r$ is given by \eqref{HR} and  $H_{\s\r}$ is as in \eqref{HSR}. We denote the spectral decomposition of $G_n$ by
\begin{equation*}
G_n = \sum_{l=1}^{\nu_n} \gamma_n(l) P_n(l),  
\end{equation*}
where $1\le \nu_n\le\dim\h_{\s,n}$ and the $\{\gamma_n(l)\}_{l=1}^{\nu_n}$ are the distinct eigenvalues of $G_n$ with associated eigenprojections  $P_n(l)$ of dimensions $\ge 1$. The eigenvalues of $G$ are
\begin{equation}
\label{specG}
{\rm spec}(G) = \big\{ F(\gamma_1(l_1),\ldots,\gamma_N(l_N))\ :\ l_j=1,\ldots,\nu_j\big\}.
\end{equation}
We will call the coupling {\it nondegenerate} if all eigenvalues of $G$ are simple. This necessitates in particular that $\dim P_j(l)=1$ for all $j$ and all $l$. As we explain below, the nondegeneracy can be viewed as a {\it generic} situation. For nondegenerate couplings each eigenvalue 
\begin{equation*}
\Gamma(l_1,\ldots,l_N) \equiv F(\gamma_1(l_1),\ldots,\gamma_N(l_N)) 
\end{equation*}
of $G$ has the associated rank one eigenprojection 
\begin{equation*}
P(l_1,\ldots,l_N) \equiv P_1(l_1)\otimes\cdots \otimes P_N(l_N),
\end{equation*}
leading to the spectral decomposition
\begin{equation*}
G = \sum_{l_1,\ldots,l_N} \Gamma(l_1,\ldots,l_N) P(l_1,\ldots,l_N).
\end{equation*}
The  density matrix of the system resulting from the nonselective measurement on $S$ implemented by the strong coupling with the reservoir is (Theorem \ref{thm1}),
\begin{equation*}
\rho_\s(0_+)\equiv \sum_{l_1,\ldots,l_N} P(l_1,\ldots,l_N) \rho_\s P(l_1,\ldots,l_N) = \sum_{l_1,\ldots,l_N} p_{l_1,\ldots,l_N} P_1(l_1)\otimes\cdots \otimes P_N(l_N),
\end{equation*}
where the $0\le p_{l_1,\ldots,l_N}\le 1$ satisfy $\sum_{l_1,\ldots,l_N} p_{l_1,\ldots,l_N}=1$ (they are the diagonal matrix elements of $\rho_\s$, therefore probabilities). Furthermore, from \eqref{HZeno},
\begin{equation}
\label{HZgen}
H_\z = \sum_{l_1,\ldots,l_N} P(l_1,\ldots,l_N) H_\s P(l_1,\ldots,l_N) = \sum_{l_1,\ldots,l_N} h_{l_1,\ldots,l_N} P_1(l_1)\otimes\cdots \otimes P_N(l_N),
\end{equation}
where $h_{l_1,\ldots,l_N}$ are the matrix elements of $H_\s$. Then
$$
e^{-it H_\z} = \sum_{l_1,\ldots,l_N} e^{-ith_{l_1,\ldots,l_N}}P(l_1,\ldots,l_N)
$$
and we obtain from Theorem \ref{thm1} that for all $t>0$,
\begin{equation}
\label{tind}
\rho_\s(t) = \sum_{l_1,\ldots,l_N} p_{l_1,\ldots,l_N} P_1(l_1)\otimes\cdots \otimes P_N(l_N).
\end{equation}
The relation \eqref{tind} shows that for nondegenerate couplings, $\rho_\s(t)$ is time independent ($t>0)$ and {\it separable}, regardless of whether the system state $\rho_\s$ before the contact with $\r$ was separable or entangled. We conclude that the ultrastrongly coupled reservoir acts as an entanglement breaking channel on $\s$. The entanglement breaking  effect is happening independently of the particular choice of the coupling $G$ of the form \eqref{HSR}, and regardless of whether $H_\s$ is a local Hamiltonian or not.
\medskip

{\it Genericness of the nondegeneracy.} 
In a sense, the nondegeneracy of the spectrum of $G$ is a generic situation, even in `homogeneous' systems. Consider for instance $N$ qubits, $\h_{\s,n}=\mathbb C^2$ for each $n$, each one coupled to the reservoir via $G_n=\sigma_x$ (Pauli $x$ operator with eigenvalues $\pm 1$). If $F$, \eqref{HSR} is a symmetric function in its $N$ variables, then of course the corresponding spectrum \eqref{specG} is degenerate. For instance, for $F(G_1,\ldots,G_N)=\sum_{n=1}^N G_n$, the eigenvalue $\Gamma=0$ of $G$ is $N\choose N/2$-fold degenerate ($N$ even). For $F(G_1,\ldots,G_N)=G_1\cdots G_N$, the degeneracies of the two eigenvalues $\pm 1$ of $G$ are even higher, equal to $2^{N-1}$. These exact symmetries leading to eigenvalue degeneracy are very special and unstable, though. Indeed, each qubit, even if being fabricated of the same material, will not generally have {\it exactly} the same energy levels (or eigenvalues of $G$), because variations naturally occur due to production imprecision or laboratory operating conditions. One may then consider that the levels of each operator $G_n$ will slightly deviate from the precise values $\pm 1$. This can be modeled by taking for $G_n$ a random matrix of the form, say,
\begin{equation*}
G_n=\sigma_x + \mu \xi_n, 
\end{equation*}
where $\mu$ is a strength parameter and the $\xi_n$ are a family of $N$ real-valued independent, identically distributed random variables with a continuous distribution, like a centered Gaussian. As mentioned above, for $\mu=0$ the eigenvalue zero of $G=\sum_{n=1}^NG_n$ for instance, and the eigenvalues $\pm 1$ of $G=G_1\cdots G_N$ are degenerate. But as soon as $\mu\neq 0$, all eigenvalues are simple, almost surely (in the sense of probability theory). 
\bigskip

{\bf Example 1.}
We close our discussion with an example showing that entanglement can be preserved if the interaction $G$ is not of the form \eqref{HSR}. Consider $\s$ to be made of two qubits and let 
\begin{equation}
\label{notG}
G= \sigma_+ \otimes \sigma_- + \sigma_- \otimes \sigma_+,
\end{equation}
where $\sigma_\pm$ are the raising and lowering operators in the $\sigma_z$ eigenbasis $\{|0\rangle, |1\rangle\}$, where  $\sigma_z|0\rangle=|0\rangle$, $\sigma_z|1\rangle = -|1\rangle$. The Bell states
\begin{equation*}
    \ket{\psi_\pm} = \frac{1}{\sqrt{2}}(\ket{00} \pm \ket{11}),\qquad  
    \ket{\phi_\pm} = \frac{1}{\sqrt{2}}(\ket{01} \pm \ket{10}) 
\end{equation*}
are the eigenvectors of $G$: $G\ket{\phi_\pm}=\pm \ket{\phi_\pm}$ and $G\ket{\psi_\pm}=0$. Therefore, the projections appearing in Theorem \ref{thm1} are 
$$ 
P_+ = \ket{\phi_+}\bra{\phi_+}, \quad P_- = \ket{\phi_-}\bra{\phi_-}, \quad P_0 = \ket{\psi_+}\bra{\psi_+} + \ket{\psi_-}\bra{\psi_-}.
$$
It is manifest that $\rho_\s(0_+)$, given by \eqref{dyns} with $t=0_+$, can be entangled. For instance the initial state $\rho_\s=|\phi_+\rangle\langle\phi_+|$ is  invariant under the projective measurement. It will therefore remain entangled after the action of the measurement.  The difference with \eqref{HSR} is that there, each subsystem involves only one operator $G_j$, while in \eqref{notG} two non-commuting ones are involved for each qubit: $\sigma_+$ and $\sigma_-$.

\subsection{Multiple reservoirs, multiple measurements}

Consider the system $\s$ coupled to two independent reservoirs $\r 1$ and $\r 2$, according to the  Hamiltonian
\begin{equation}
\label{mr1}
    H= H_\s + H_{\r 1} + \lambda_1 G_1 \otimes \varphi_1(g_1) + H_{\r 2} + \lambda_2 G_2 \otimes \varphi_2(g_2)
\end{equation}
acting on the Hilbert space 
$$
\h = \h_\s\otimes\h_{\r 1}\otimes\h_{\r 2}.
$$
It is understood in the notation that $H_{\r 1}, \varphi_1$ and $H_{\r 2}, \varphi_2$ are observables pertaining only to the first and second reservoir, respectively, while $H_\s, G_1$ and $G_2$ are hermitian operators on $\h_\s$. 

Formally, one may understand \eqref{mr1} to be of the form \eqref{Ham}, with the pair $\s$ and $\r 1$ making a new `system'. By keeping $\lambda_1$ fixed and taking $|\lambda_2|\rightarrow\infty$, Theorem \ref{thm1} would then show that the dynamics of $\s$ plus $\r 1$ is given by a nonselective measurement relative to the spectral projections $P_l$ of $G_2$ followed by a Zeno dynamics with Hamiltonian
\begin{equation}
\label{zeno2res}
H_{\rm Z} =\sum_{l=1}^\nu P_l \big( H_\s + H_{\r1} + \lambda_1 G_1 \otimes \varphi_1(g_1) \big) P_l .
\end{equation}
The caveat is that Theorem \ref{thm1} was shown for a finite dimensional $\h_\s$, so it is not immediately applicable to this situation. Nevertheless, we now demonstrate that this result is correct, provided the state $\omega_{\r 1}$ satisfies the following regularity condition: For all $n \in \mathbb{N}$  we have 
\begin{equation}
\label{r1regularA}
\max_{1 \leq j \leq n} \; \sup_{0\le t_1,t_2, \ldots, t_j\le t} \Big| \omega_{\r1} \Big(\varphi_1(e^{i\omega t_1} g_1) \cdots \varphi_1(e^{i\omega t_j}g_1) \Big) \Big| \leq B_n,
\end{equation} 
for some numbers $B_n\ge0$ satisfying 
\begin{equation}\label{r1regularB}
    \sum_{n \geq 0} \frac{\alpha^n}{n!} B_n < \infty \qquad \mbox{for any $\alpha>0$.}
\end{equation}

The conditions \eqref{r1regularA} and \eqref{r1regularB} hold in particular for quasifree states $\omega_\r$ for which Wick's theorem applies (such as equilibrium states at any temperature). Those states satisfy for $n\in\mathbb N$ \cite{BR}
\begin{equation*}
\omega_\r\big( \varphi(f_1)\cdots \varphi(f_{2n})\big) = \sum_{p\in\mathcal P} \prod_{(p_1,p_2)\in p} \omega_\r\big(\varphi(f_{p_1})\varphi(f_{p_2}) \big),
\end{equation*}
where the $f_1,\ldots, f_{2n}$ are arbitrary functions in $L^2(\mathbb R^3, d^3k)$ (and odd moments vanish). The number of pairings $p$ is $|\mathcal P|=\frac{(2n)!}{2^n n!}$ and by using $|\omega_\r(\varphi(f)\varphi(g))|\le C$ one readily verifies that  \eqref{r1regularA} and \eqref{r1regularB} hold. This gives a rich and physically relevant class of reservoir states we can treat.

\begin{thm}
\label{thm2res}
Assume the condition \eqref{r1regularA}, \eqref{r1regularB} and let $P_l$, $l=1,\ldots,\nu$ be the spectral projections of $G_2$. 
Then, for any observable $A$ of the system and any $t>0$  we have 
\begin{equation}
\label{lim2res}
    \lim_{|\lambda_2| \to \infty} \omega_\s \otimes \omega_{\r1} \otimes \omega_{\r2} \Big( e^{i tH } (A \otimes \bbbone_{\r1} \otimes \bbbone_{\r2}) e^{-it H} \Big) = \omega_\s \otimes \omega_{\r1} \Big( e^{i tH_\z} \big(\sum_{l=1}^\nu P_l A P_l \otimes \bbbone_{\r1}\big) e^{-i tH_\z }\Big) ,
\end{equation}
where $H_\z$ is the Zeno Hamiltonian \eqref{zeno2res}.
\end{thm}

We present a proof of Theorem \ref{thm2res} in Section \ref{sect:proof2res}. Note that the system part of the coupling operator in $H_\z$, \eqref{zeno2res} {\it after} the ultrastrong coupling interaction $|\lambda_2|\rightarrow\infty$ is 
\begin{equation}
\label{G'}
G'_1 = \sum_{l=1}^\nu P_l G_1P_l.
\end{equation}
Therefore, if we perform the $|\lambda_1|\rightarrow\infty$ limit (after $|\lambda_2|\rightarrow\infty$), then the resulting nonselective measurement of $
\s$ is performed according to the observable $G'_1$ which commutes with $G_2$. As a consequence, by having $\s$ in contact with several reservoirs and sequentially taking ultrastrong coupling limits to the different reservoirs, one cannot implement sequential nonselective measurements associated to incompatible (not commuting) observables. This stems from the fact that the first ultrastrong coupling affects the interaction operators of all following coupling processes.

If one did want to implement successive nonselective measurements of non-commuting system observables $G_1,\ldots,G_m$, one would need to couple the system $\s$ to $m$ individual environments in a successive manner, one at the time. This is the {\it collision model} setup, where $\s$ interacts (`collides') with $\r 1$ alone first, then is decoupled from $\r 1$ and collides with $\r 2$ alone, and so on. The resulting system state after the first collision is given by \eqref{dyns} (take $H_\s=0$) $\rho_\s^{(1)} = \sum_{l_1} P^{(1)}_{l_1}\rho_\s P^{(1)}_{l_1}$. This is the initial state for the second collision, after which the system state is $\rho_\s^{(2)}=\sum_{l_1,l_2} P^{(1)}_{l_2}P^{(1)}_{l_1} \rho_\s P_{l_1}^{(1)} P_{l_2}^{(2)}$. After $m$ collisions, the system is in the state
\begin{equation*}
\rho_\s^{(m)} = \sum_{l_1,\ldots,l_m} P^{(m)}_{l_m} \cdots P^{(1)}_{l_1} \rho_\s   P^{(1)}_{l_1}\cdots P^{(m)}_{l_m},
\end{equation*}
where $P_l^{(j)}$ are the spectral projections of $G_j$, the operator describing the interaction of $\s$ with the $j$th reservoir (as per \eqref{Ham}). 
\medskip

Let us finally discuss the system density matrix  $\rho_\s(t)$ associated to \eqref{lim2res}, which is defined by setting ${\rm tr}_\s(\rho_\s(t) A)$ equal to \eqref{lim2res} for all $A$. Suppose that all the projections are rank one, $P_l=|\psi_l\rangle\langle\psi_l|$, that is, all the eigenvalues of $G_2$ are simple, then
\begin{equation}\label{rank1}
    e^{-it H_\z} = \sum_{l} P_l\otimes e^{-it (h_l +H_{\r1} +\lambda_1 \eta_l \varphi(g_1))},
\end{equation}
where $\eta_l=\langle\psi_l, G_1\psi_l\rangle$ and $h_l = \langle \psi_l,H_\s\psi_l\rangle$. Substituting \eqref{rank1} into \eqref{lim2res} we see that the system density matrix is constant in time after the first measurement and is in a state which only depends on the interaction with $\r 2$. {\it The system is entirely decoupled from $\r 1$ by the ultrastrong coupling to $\r 2$.} This happens if all measurement projections (associated to $G_2$) have rank one. {\it However}, a $G_2$ with degenerate spectrum does not in general decouple $\s$ from $\r 1$ and leads to a rich, usually non-Markovian, dynamics of $\s$. We illustrate this with an example.
\medskip

{\bf Example 2.}
Take a two qubit system interacting with  two reservoirs $\r1$ and $\r2$ as per \eqref{mr1}, with a coupling 
$$
G_2= \sigma_z \otimes \bbbone_2 + \bbbone_1 \otimes \sigma_z. 
$$
Here, $\bbbone_1,\bbbone_2$ are the identities on the first and second qubit, respectively (so $G_2$ acts on both qubits). 
The spectral projections of $G_2$ are $P_+ = \ket{00}\bra{00}$, $P_- = \ket{11}\bra{11}$ (both rank one) and $P_0= \ket{01}\bra{01} + \ket{10}\bra{10}$ (rank two). A nontrivial evolution will generally take place in the two dimensional Zeno subspace $\h_{\rm eff}:={\rm Ran} P_0$, which can be identified as the state space of an {\it effective single qubit}. The dynamics generated by the Zeno Hamiltonian leaves $\h_{\rm eff}$ invariant | leading to a dynamics of the effective qubit. Namely, let
\begin{equation}
\label{effqbit}
\rho_\s = \frac{1+z}{2} \ket{01}\bra{01} + \frac{1-z}{2} \ket{10}\bra{10} + \frac{c}{2} \ket{01}\bra{10} + \frac{c^*}{2} \ket{10}\bra{01}
\end{equation}
be an initial density matrix, with $-1\leq z \leq 1$ and $ 0 \leq |c|^2 \leq 1-z^2 $.
After the measurement and Zeno evolution to time $t>0$, the effective qubit state is again of the form \eqref{effqbit}, with time dependent $z(t)$ and $c(t)$. For instance, if 
$$
H_\s=G_1= \varepsilon_1 \sigma_z\otimes \bbbone_2 +  \bbbone_1 \otimes \varepsilon_2\sigma_z 
$$
we find that $z(t)=z$ (constant populations), while the coherence is
\begin{equation*}
c(t)  = e^{2it \Delta }  D(t) c , \qquad D(t) = \omega_{\r 1}\Big( W\big(\lambda_1 \Delta \tfrac{e^{i\omega t}-1}{i\omega}g_1\big)\Big),
\end{equation*}
with $\Delta=\varepsilon_1-\varepsilon_2$. This is the well known pure dephasing model. If $\omega_{\r 1}$ is in thermal equilibrium at inverse temperature $\beta$, the decoherence function $D(t)$ has the explicit expression,
$$
D_\beta(t) = \exp\Big[-\lambda^2_1 \Delta^2 \int_{\mathbb R^3} |g_1(k)|^2 \coth(\beta\omega(k)/2) \frac{\sin^2(\omega(k) t/2)}{\omega^2(k)} d^3k\Big].
$$
Depending on the explicit form of $g(k)$ and on the value of $\beta$, the decoherence function $D_\beta(t)$ may be non-monotonic in $t$, thus describing a non-Markovian quantum evolution \cite{HJM13}. In this respect the system dynamics resulting from the ultrastrong coupling in the presence of several reservoirs is richer than for a single reservoir  | the latter is always Markovian because it is the composition of a projective measurement and a unitary evolution. By choosing appropriate $H_\s$ and $G_1$ one will get dissipative qubit evolutions where populations are modified as well.

\section{Proof of the results}

The proof of Theorem \ref{thm2} (which implies Theorem \ref{thm1}) is the main technical part of the paper. The proof of Theorem \ref{thm2res} is a variation which  takes into account the infinite dimensional nature of the reduced system $\s+\r1$ and the unboundedness of the resulting Zeno Hamiltonian.

\subsection{Proof of Theorem \ref{thm2}}
\label{sec:proofthm2}

 We set
\begin{equation}
\label{K}
K=K_\lambda = H_\r +\lambda G\otimes \varphi(g).
\end{equation}
The Dyson series gives 
\begin{equation}
\label{D}
e^{i t H} e^{-i t K} = \bbbone +\mathcal D,\qquad 
\mathcal D = \sum_{n\ge 1} i^n \int_{0\le t_n\le\cdots\le t_1\le t} H_\s(t_n)\cdots H_\s(t_1), 
\end{equation}
where
$$
H_\s(t) = e^{i t K}H_\s e^{-itK}.
$$
The series converges for all $t\ge 0$, and it does so uniformly in $\lambda\in\mathbb R$. Our goal is to analyze (see \eqref{thmeqn}) the large $\lambda$ limit of $\omega_\s\otimes\omega_\r(e^{itH}\{B(\lambda)\otimes W(h)\}e^{-itH})$, where
\begin{equation}
\label{Blambda}
B(\lambda) = e^{\frac i2\lambda G {\rm Re}\langle g, \frac{1-e^{i\omega t}}{\omega} h\rangle} Ae^{\frac i2\lambda G {\rm Re}\langle g, \frac{1-e^{i\omega t}}{\omega} h\rangle}.
\end{equation}
Using \eqref{D} we have 
\begin{eqnarray}
\lefteqn{\omega_\s\otimes\omega_\r \big(e^{it H} \big(B(\lambda)\otimes W(h)\big) e^{-itH}\big) }\nonumber\\
 &=& \omega_\s\otimes\omega_\r\big(e^{it K} \big(B(\lambda)\otimes W(h)\big) e^{-itK}\big) \nonumber\\
 && +\omega_\s\otimes\omega_\r\big(\mathcal D e^{it K} \big(B(\lambda)\otimes W(h)\big) e^{-itK}\big) \nonumber\\
& & +\omega_\s\otimes\omega_\r\big( e^{it K} \big(B(\lambda)\otimes W(h)\big) e^{-itK}\mathcal D^* \big) \nonumber\\
& & +\omega_\s\otimes\omega_\r\big(\mathcal D e^{it K} \big(B(\lambda)\otimes W(h)\big) e^{-itK}\mathcal D^*\big)
 \nonumber\\
&\equiv & T_1+T_2+T_3+T_4.
 \label{sum}
\end{eqnarray} 
We first calculate the limit $|\lambda|\rightarrow\infty$ of $T_1$. To do this, we use the following result.

\begin{lem}
\label{lem:1}
For $l=1,\ldots,\nu$, let $f_l(k)$ be such that 
\begin{equation}
    \label{fl}
i\omega(k) f_l(k)=\lambda\gamma_l g(k)
\end{equation}
or for short, $i\omega f_l=\lambda\gamma_l g$.
Then we have for any $A\in\mathcal B(\h_\s)$ and any  $h\in L^2(\mathbb R^3,d^3k)$,
\begin{align}
e^{it K} (A\otimes W(h)) e^{-itK}   =  & \sum_{l,r=1}^\nu P_l A P_r \, e^{-\frac i2 {\rm Im} \langle (e^{i\omega t} -1 -i\omega t) (f_l-f_r),(f_l+f_r)\rangle}  e^{-\frac i2{\rm Im} \langle (f_l+f_r), (1-e^{i\omega t}) h\rangle}\nonumber\\
& \times W\big((e^{i\omega t}-1)(f_l-f_r) +e^{i\omega t}h\big).
\label{eq1lem1}
\end{align}
In particular, with $B(\lambda)$ given as in \eqref{Blambda}, we have
\begin{align}
e^{it K} (B(\lambda)\otimes W(h)) e^{-itK} = &  \sum_{l,r=1}^\nu P_l A P_r \, e^{-\frac i2 \lambda^2(\gamma^2_l-\gamma^2_r){\rm Im} \langle g, \frac{e^{-i\omega t}+it\omega}{\omega^2}g\rangle}  \nonumber\\
&  \ \times W\big((e^{i\omega t}-1)(f_l-f_r)+e^{i\omega t}h\big).
\label{eq2lem1}
\end{align}
\end{lem}

\medskip

{\it Remark.} If $\omega(k)=0$ then in the expression \eqref{eq1lem1} we have $(e^{i\omega(k) t}-1)(f_l(k)-f_r(k))= t\lambda(\gamma_l-\gamma_r)g(k)$, see also \eqref{fl}. 

\medskip

\noindent
{\bf Proof of Lemma \ref{lem:1}.} We propose two proofs, one based on the polaron transformation, the other based on the the Trotter product formula. While the first one might seem a bit shorter it uses a condition on the infrared behaviour of the form factor $g$ which is in actual fact not needed. The proof based on the Trotter product formula works without this condition.

{\it Proof based on the polaron transformation. } The following relations are well known,
\begin{eqnarray*}
W(f) H_\r  W(f)^* &=& H_\r-\varphi(i\omega f) +\tfrac12 \|\sqrt\omega f\|_{L^2}^2\\
W(f) \varphi(g) W(f)^* &=& \varphi(g) - {\rm Im}\langle f,g\rangle.
\end{eqnarray*}
Here, the $L^2\equiv L^2(\mathbb R^3,d^3k)$ norm of a (complex valued) function $F(k)$, $k\in\mathbb R^3$, is given by
$$
\| F\|_{L^2} = \sqrt{\int_{\mathbb R^3}|F(k)|^2 d^3k}.
$$
Setting (see \eqref{fl}) $i\omega f=\lambda\gamma g$ we get 
$$
W(f) \big(H_\r +\lambda \gamma\varphi(g)\big) W(f)^* = H_\r-\tfrac12 \|\sqrt\omega f\|^2_{L^2} = H_\r -\tfrac12 \lambda^2\gamma^2 \|g/\sqrt\omega\|_{L^2}^2,
$$
and so,
$$
e^{it K}=\sum_{l=1}^\nu P_l\, e^{it(H_\r+\lambda \gamma_l\varphi(g))} =\sum_{l=1}^\nu P_l\, W(f_l)^* e^{it H_\r} W(f_l) e^{-\frac{it}{2}\| \sqrt{\omega} f_l\|_{L^2}^2}.
$$
This approach assumes that $\frac{1}{\sqrt\omega}g\in L^2({\mathbb R^3},d^3k)$. 
We have
\begin{align*}
e^{itK}(A\otimes W(h))e^{-itK}  = & \sum_{l,r=1}^\nu P_l A P_r \, e^{-\frac{it}{2} (\|\sqrt\omega f_l\|_{L^2}^2 - \|\sqrt\omega f_r\|_{L^2}^2)} \nonumber\\
& \times W(f_l)^*e^{it H_\r} W(f_l) \ W(h)\ W(f_r)^* e^{-it H_\r}W(f_r)\nonumber\\
= & \sum_{l,r=1}^\nu P_l A P_r \, e^{-\frac i2 {\rm Im} \big\langle (e^{i\omega t} -1 -i\omega t)(f_l-f_r),(f_l+f_r)\big\rangle}  e^{-\frac i2 {\rm Im} \big\langle (f_l+f_r), (1-e^{i\omega t}) h\big\rangle}\nonumber\\
& \times W\big((e^{i\omega t}-1)(f_l-f_r)+e^{i\omega t}h\big),
\end{align*}
where we used $W(f)^*=W(-f)$, the CCR (canonical commutation relations) and the Bogolyubov dynamics,
\begin{equation}
\label{used}
\mbox{$W(f)W(g)=e^{-\frac i2{\rm Im}\langle f,g\rangle}W(f+g)$ \ and \ $e^{it H_\r}W(f)e^{-itH_\r} = W(e^{it \omega}f)$}.
\end{equation}
This shows \eqref{eq1lem1} with the proviso that $f_l$ defined by \eqref{fl} is square integrable. It is apparent, though, that the possible singularity introduced by the factor $1/\omega$ in $f_l$ is compensated by the term $e^{i\omega t}-1$ in \eqref{eq1lem1}, so in actual fact the result \eqref{eq1lem1} holds under the sole condition that $g\in L^2({\mathbb R}^3,d^3k)$, as we show now.

{\it Proof based on the Trotter product formula.} We diagonalize $G$,
\begin{equation*}
e^{itK} \big(A\otimes W(h)\big)e^{-itK} = \sum_{l,r}P_l AP_r\otimes e^{it(H_\r+\lambda\gamma_l \varphi(g))} W(h)  e^{-it(H_\r+\lambda\gamma_r \varphi(g))}.
\end{equation*}
For brevity of the notation, we shall absorb the constant $\lambda$ into the form factor $g$ and put it back at the end of the calculation. 
By the Trotter product formula \cite{Zagrebnov},
\begin{equation*}
e^{it(H_\r+\gamma_l \varphi(g))} W(h)  e^{-it(H_\r+\gamma_r \varphi(g))} = \lim_{n\rightarrow\infty} \Big(e^{i\frac tn H_\r} W(\tfrac{t}{n}\gamma_lg\big)\Big)^n W(h) \Big(W(-\tfrac{t}{n}\gamma_r g\big)e^{-i\frac tn H_\r} \Big)^n,
\end{equation*}
where $W(\tfrac{t}{n}\gamma g)=e^{i\frac tn\gamma\varphi(g)}$. Setting $P(l,n)=e^{i\frac tn H_\r} W(\tfrac{t}{n}\gamma_lg\big)$ and $P(r,n)=W(-\tfrac{t}{n}\gamma_r g\big)e^{-i\frac tn H_\r}$ we find
\begin{eqnarray}
P(l,n)^n W(h)P(r,n)^n &=& P(l,n)^{n-1}\Big( e^{i\frac tn H_\r} W(\tfrac{t}{n}\gamma_lg\big) W(h)W(-\tfrac{t}{n}\gamma_r g\big)e^{-i\frac tn H_\r}\Big)P(r,n)^{n-1}\nonumber\\
&=& e^{-\frac i2 \Phi_1(n)} P(l,n)^{n-1} W\big(h_1(n)\big) P(r,n)^{n-1},
\label{recur}
\end{eqnarray}
where
\begin{equation*}
 \Phi_1(n,t) ={\rm Im} \big\langle \tfrac tn(\gamma_l+\gamma_r)g,h\big\rangle,\qquad h_1(n) = e^{i\omega \frac{t}{n}}\big(\tfrac{t}{n} (\gamma_l-\gamma_r)g+h\big).
\end{equation*}
To arrive at \eqref{recur}, we used the CCR to combine the product of the three Weyl operators into a single one, thus producing a phase $\Phi_1$, and we used that $H_\r$ implements the dynamics of the reservoir, as in \eqref{used}. We continue the process to obtain, for $1\le k\le n$,
$$
P(l,n)^nW(h) P(r,n)^n = e^{-\frac i2\Phi_k(n)} P(l,n)^{n-k} W\big(h_k(n)\big) P(r,n)^{n-k}
$$
with explicit formulas for $\Phi_k$ and $h_k$, see {\it e.g.} the proof of Proposition 7.4 in \cite{MSB2008}. Doing this $k=n$ times and taking $n\rightarrow\infty$ gives 
$$
\lim_{n\rightarrow\infty} P(l,n)^n W(h) P(r,n)^n = \,   e^{-\frac i2 \Phi} W\big((e^{i\omega t}-1)(f_l-f_r)+e^{i\omega t}h\big)
$$
with $\Phi =(\gamma_l^2-\gamma_r^2) {\rm Im} \langle \frac{e^{i\omega t} -1 -i\omega t}{\omega^2}g,g \rangle  +(\gamma_l+\gamma_r) {\rm 
Re } \langle g, \frac{1-e^{i\omega t}}{\omega} h\rangle$. Remembering that $g$ was actually $\lambda g$ we recover the expression \eqref{eq1lem1} without having assumed that $f_l$ is square integrable. 

The relation \eqref{eq2lem1} follows  from \eqref{eq1lem1} and \eqref{fl}. The proof of Lemma \ref{lem:1} is complete. \hfill $\qed$

\medskip
\noindent
Let us now treat each term $T_j$ in \eqref{sum}.

$\bullet$ We first take the limit $|\lambda|\rightarrow\infty$ of $T_1$ (see \eqref{sum}).  We have from Lemma \ref{lem:1},
\begin{align}
T_1 = & \omega_\s\otimes\omega_\r\big(e^{it K} \big(B(\lambda)\otimes W(h)\big) e^{-itK}\big) \nonumber\\
  = &  \sum_{l,r=1}^\nu \omega_\s\big( P_l A P_r\big) \, e^{-\frac i2 \lambda^2(\gamma^2_l-\gamma^2_r){\rm Im} \langle g, \frac{e^{-i\omega t}+it\omega}{\omega^2}g\rangle} 
   \omega_\r\Big( W\big((e^{i\omega t}-1)(f_l-f_r)+e^{i\omega t}h\big)\Big).
\label{term1}
\end{align}
The expectation of the Weyl operator is (recall \eqref{gaus} and also the notation \eqref{fl})
\begin{equation}
\omega_\r\Big( W\big((e^{i\omega t}-1)(f_l-f_r)+e^{i\omega t}h\big)\Big) = e^{-\frac14 \big\| \mathcal C^{1/2} \big[\lambda(\gamma_l-\gamma_r)\frac{e^{i\omega t}-1}{i\omega}g+e^{i\omega t}h \big]\big\|_{L^2}^2},
\label{resexp}
\end{equation}
where $\mathcal C\ge \bbbone$ is the covariance operator. Now 
\begin{eqnarray*}
\left\| \mathcal C^{1/2}\big[\lambda(\gamma_l-\gamma_r)\frac{e^{i \omega t}-1}{i\omega}g+e^{i\omega t}h\big]\right\|_{L^2} 
  &\ge& \left\| \lambda(\gamma_l-\gamma_r)\frac{e^{i \omega t}-1}{i\omega}g+e^{i\omega t}h \right\|_{L^2} \nonumber\\
&\ge&  |\lambda| \, |\gamma_l-\gamma_r|\,  \left\| \frac{e^{i\omega t}-1}{i\omega}g\right\|_{L^2} - \big\|h\big\|_{L^2}\nonumber\\
&\ge& |\lambda| \, |\gamma_r-\gamma_l| \, \mu -\| h\|_{L^2}
\end{eqnarray*}
where $\mu$ may depend on $t$ and satisfies $\mu>0$ for all $t>0$, since the $L^2$ norm does not vanish. It follows that for any $t>0$, and $\gamma_l\neq\gamma_r$,
$$
\lim_{|\lambda|\rightarrow\infty} \big\| \mathcal C^{1/2}\big[\lambda(\gamma_l-\gamma_r)\frac{e^{i\omega t}-1}{i\omega}g+e^{i\omega t}h\big]\big\|_{L^2}^2 =\infty,
$$
so the limit $|\lambda|\rightarrow\infty$ of \eqref{resexp} vanishes whenever $t>0$ and $\gamma_l\neq\gamma_r$. Therefore, from \eqref{term1},
\begin{equation}
\label{T1}
\lim_{|\lambda|\rightarrow\infty}  T_1= \sum_{l=1}^\nu \omega_\s(P_lAP_l) \omega_\r\big(W(e^{i\omega t}h)\big),\qquad t>0.
\end{equation}

\medskip

$\bullet$ We take the limit $|\lambda|\rightarrow\infty$ of $T_2$ (see \eqref{sum}). Using \eqref{eq2lem1} and \eqref{D} we have 
\begin{align}
T_2 = & \ \omega_\s\otimes\omega_\r\big(\mathcal D e^{it K} \big(B(\lambda)\otimes W(h)\big) e^{-itK}\big)\nonumber\\
= &  \sum_{l,r=1}^\nu \, e^{-\frac i2 \lambda^2(\gamma^2_l-\gamma^2_r){\rm Im} \langle g, \frac{e^{-i\omega t}+it\omega}{\omega^2}g\rangle} 
\sum_{n\ge 1} i^n \int_{0\le t_n\le\cdots\le t_1\le t}\nonumber\\
 & \times \omega_\s\otimes\omega_\r\Big(H_\s(t_n)\cdots H_\s(t_1) P_lAP_r\   W\big((e^{i\omega t}-1)(f_l-f_r)+e^{i\omega t}h\big)\Big).
 \label{term2}
\end{align}
Next, by Lemma \ref{lem:1},
\begin{align}
H_\s(t_n)\cdots H_\s(t_1)  = & \sum_{l_0,\ldots,l_n=1}^\nu P_{l_n}H_\s P_{l_{n-1}}\cdots P_{l_1}H_\s P_{l_0}
  e^{-\frac i2 \lambda^2\sum_{j=1}^n(\gamma^2_{l_j}-\gamma^2_{l_{j-1}}){\rm Im} \langle g, \frac{e^{-i\omega t_j}+ it_j\omega}{\omega^2}g\rangle} \nonumber\\
  & \times W\big((e^{it_n\omega}-1)(f_{l_n}-f_{l_{n-1}})\big) \cdots W\big((e^{it_1\omega}-1)(f_{l_1}-f_{l_0})\big)\nonumber\\
   = & \sum_{l_0,\ldots,l_n=1}^\nu P_{l_n}H_\s P_{l_{n-1}}\cdots P_{l_1}H_\s P_{l_0}
  e^{-\frac i2\lambda^2 \Phi_1} W\Big( \sum_{j=1}^n (e^{it_j\omega}-1)(f_{l_j}-f_{l_{j-1}}) \Big),
\label{prodHs}
\end{align}
where the phase is
\begin{align}
\Phi_1  =& \ \Phi_1(t_1,\ldots,t_n,l_0,\ldots,l_n) \nonumber\\
 = &\sum_{j=1}^n(\gamma^2_{l_j}-\gamma^2_{l_{j-1}}){\rm Im} \left\langle g, \frac{e^{-i\omega t_j}+it_j\omega}{\omega^2}g \right\rangle \nonumber\\
& -  \sum_{r=1}^{n-1} \sum_{j=1}^r (\gamma_{l_j}-\gamma_{l_{j-1}}) (\gamma_{l_{r+1}}-\gamma_{l_r}) {\rm Im }\left\langle g, \frac{(1-e^{-it_j\omega})(1-e^{it_{r+1}\omega})}{\omega^2} g \right\rangle.
\label{phi1}
\end{align}
The double sum in \eqref{phi1} comes from the commutation relations when combining the product of the Weyl operators into a single Weyl operator. 
We insert \eqref{prodHs} into \eqref{term2},
\begin{align}
T_2 = & \ \omega_\s\otimes\omega_\r\big(\mathcal D e^{it K} \big(B(\lambda)\otimes W(h)\big) e^{-itK}\big)\nonumber\\
= &  \sum_{l_0,\ldots,l_n, r_0=1}^\nu \, \sum_{n\ge 1} i^n \int_{0\le t_n\le\cdots\le t_1\le t} e^{-\frac i2 \Phi_2} \omega_\s\big(P_{l_n}H_\s P_{l_{n-1}}\cdots P_{l_1}H_\s P_{l_0} A P_{r_0}\big)\nonumber\\
 &  \times \omega_\r \Big(W\big( \sum_{j=1}^n (e^{it_j\omega}-1)(f_{l_j}-f_{l_{j-1}}) \big) W\big((e^{it\omega}-1)(f_{l_0}-f_{r_0})+e^{i\omega t}h\big)\Big),
 \label{term2eq2}
\end{align}
where 
$$
\Phi_2 = \lambda^2 \Phi_1 +  \lambda^2 (\gamma^2_{l_0}-\gamma^2_{r_0}){\rm Im} \left\langle g, \frac{e^{-i\omega t}+it\omega}{\omega^2}g \right\rangle. 
$$
Now we evaluate the average over the Weyl operators in \eqref{term2eq2},
\begin{align}
\omega_\r \Big(W\big( \sum_{j=1}^n (e^{it_j\omega}-1)&(f_{l_j}-f_{l_{j-1}}) \big) W\big((e^{it\omega}-1)(f_{l_0}-f_{r_0})+e^{i\omega t}h\big)\Big)\nonumber\\
 = & \ e^{-\frac i2{\rm Im}\sum_{j=1}^n\langle (e^{it_j\omega}-1)(f_{l_j}-f_{l_{j-1}}),(e^{it\omega}-1)(f_{l_0}-f_{r_0}) +e^{i\omega t}h\rangle}\nonumber\\
 &\times \omega_\r \Big(W\Big( \sum_{j=1}^n (e^{it_j\omega}-1)(f_{l_j}-f_{l_{j-1}})  + (e^{it\omega}-1)(f_{l_0}-f_{r_0})+e^{i\omega t}h\Big)\Big)\nonumber\\
=& \ e^{-\frac i2{\rm Im}\sum_{j=1}^n\langle \frac{e^{it_j\omega}-1}{i\omega}\lambda(\gamma_{l_j}-\gamma_{l_{j-1}}) g, \frac{e^{it\omega}-1}{i\omega}\lambda (\gamma_{l_0}-\gamma_{l_{r_0}})g +e^{i\omega t}h\rangle}\nonumber\\
& \times e^{-\frac14 \big\| \mathcal C^{1/2} \big[ \sum_{j=1}^n \frac{e^{it_j\omega}-1}{i\omega}\lambda (\gamma_{l_j}-\gamma_{l_{j-1}})g  + \frac{e^{it\omega}-1}{i\omega}\lambda  (\gamma_{l_0}-\gamma_{r_0})g+e^{i\omega t}h\big] \big\|^2_{L^2}}.  
\label{bigWeyl}
\end{align}

\begin{lem}
\label{lem:2}
Suppose $g(k_0)\neq 0$ for some $k_0\in\mathbb R^3$ and that $g$ is continuous in a neighbourhood $\mathcal N_0$ of $k_0$. Suppose that either $\omega(k_0)=0$ for some  $k_0\in\mathcal N_0$ or that $\omega(\mathcal N_0)$ contains two points $\omega_1,\omega_2$ such that $\omega_2/\omega_1\not\in\mathbb Q$. Let $\gamma_{r_0}, \gamma_{l_0}, \gamma_{l_1},\ldots,\gamma_{l_n}$ be given, not all equal, and let $t>0$ be given. Then we have 
\begin{equation}
\label{eqlem2}
\lim_{|\lambda|\rightarrow\infty } e^{-\frac14 \big\| \mathcal C^{1/2} \big[ \sum_{j=1}^n \frac{e^{it_j\omega}-1}{i\omega}\lambda (\gamma_{l_j}-\gamma_{l_{j-1}})g  + \frac{e^{it\omega}-1}{i\omega}\lambda  (\gamma_{l_0}-\gamma_{r_0})g+e^{i\omega t}h\big] \big\|^2_{L^2}} =0
\end{equation}
for almost every $(t_1,\ldots,t_n)$ belonging to $\mathbb R^n$.
\end{lem}

Lemma \ref{lem:2} shows that the limit does not vanish only if all the indices $\gamma$ have the same value ($t\neq 0$).
\medskip

{\it Proof of Lemma \ref{lem:2}. } 
We have 
\begin{align}
\big\| \mathcal C^{1/2}& \big[ \sum_{j=1}^n \frac{e^{it_j\omega}-1}{i\omega}\lambda (\gamma_{l_j}-\gamma_{l_{j-1}})g  + \frac{e^{it\omega}-1}{i\omega}\lambda  (\gamma_{l_0}-\gamma_{r_0})g+e^{i\omega t}h\big] \big\|_{L^2} \nonumber\\
&\ge \big\|   \sum_{j=1}^n \frac{e^{it_j\omega}-1}{i\omega}\lambda (\gamma_{l_j}-\gamma_{l_{j-1}})g  + \frac{e^{it\omega}-1}{i\omega}\lambda  (\gamma_{l_0}-\gamma_{r_0})g+e^{i\omega t}h  \big\|_{L^2} \nonumber\\
 &\ge |\lambda| \ \big\| \sum_{j=1}^n \frac{e^{it_j\omega}-1}{i\omega} (\gamma_{l_j}-\gamma_{l_{j-1}})g  + \frac{e^{it\omega}-1}{i\omega}  (\gamma_{l_0}-\gamma_{r_0})g \big\|_{L^2} -\|h\|_{L^2}.
 \label{normlem2}
\end{align}
We are going to show that the $L^2$ norm of first term on the right side of \eqref{normlem2} is strictly positive unless all the $\gamma$ are the same. Then \eqref{eqlem2} follows. As $g(k_0)\neq 0$ and $g$ is continuous in a neighbourhood of $k_0$, we can assume without loss of generality that $g(k_0)\neq 0$ for all $k\in\mathcal N_0$. Then the equality
\begin{align}
\big\| \sum_{j=1}^n \frac{e^{it_j\omega}-1}{i\omega} (\gamma_{l_j}-\gamma_{l_{j-1}})g  + \frac{e^{it\omega}-1}{i\omega}  (\gamma_{l_0}-\gamma_{r_0})g \big\|_{L^2}=0 
\label{normzero}
\end{align}
implies that
\begin{equation}
\label{48}
\sum_{j=1}^n \frac{e^{it_j\omega(k)}-1}{i\omega(k)}(\gamma_{l_j}-\gamma_{l_{j-1}}) + \frac{e^{it\omega(k)}-1}{i\omega(k)}(\gamma_{l_0}-\gamma_{r_0}) =0,\qquad k\in\mathcal N_0.
\end{equation}
If $\omega(k_0)=0$ for some $k_0\in\mathcal N_0$ then since for $\omega=0$ we have the convention $\frac{e^{is\omega}-1}{i\omega}=s$ (see also after \eqref{eq2lem1}), the equation \eqref{48} for $k=k_0$ gives
\begin{equation}
\label{49}
\sum_{j=1}^n t_j\delta_j=\alpha,
\end{equation}
where $\delta_j=\gamma_{l_j}-\gamma_{l_{j-1}}$, $1\le j\le n$ and  $\alpha =-t(\gamma_{l_0}-\gamma_{r_0})$. For a given $\vec\delta=(\delta_1,\ldots,
\delta_n)^T\in\mathbb R^n$ and $\alpha\in\mathbb R$, the vectors $\vec t=(t_1,\ldots,t_n)^T\in \mathbb R^n$ satisfying \eqref{49} span a plane with normal $\vec\delta$ and an offset from the origin determined by $\alpha$. Therefore, those $\vec t$ belong to a set of measure zero in $\mathbb R^n$. We conclude that \eqref{48} can hold for a set of $\vec t$ of positive measure only if $\vec\delta=0$, and hence $\alpha=0$, and hence $\gamma_{l_0}=\gamma_{r_0}$ (for $t\neq 0$). This shows Lemma \ref{lem:2} in case $\omega(k_0)=0$ for some $k_0\in\mathcal N_0$.   

If $\omega(k_1)\equiv\omega_1\neq 0$ for some $k_1\in\mathcal N_0$, then multiplying \eqref{48} for $k=k_1$ by $i\omega_1$ and taking the imaginary part yields
\begin{equation}
\label{49.1}
\sum_{j=0}^n v_j \, \delta_j=\alpha,
\end{equation}
where the $\delta_j$ are as above and where $v_j=\sin(\omega_1 t_j)$ and $\alpha = -\sin(\omega_1t)(\gamma_{l_0}-\gamma_{r_0})$. As above if $\vec\delta$ is not zero this holds for $\vec v=(v_1,\ldots,v_n)^T$ in a set of measure zero in $\mathbb R^n$. The values of $\vec t$ for which this holds is then also a set of measure zero in $\mathbb R^n$ for the following reason. The Jacobian of the transformation 
$$
F:\mathbb R^n\rightarrow\mathbb R^n, \qquad \vec v = F(\vec t\,)
$$
is
$$
DF(\vec t\,) = \omega_1^n\prod_{j=1}^n \cos(\omega_1 t_j).
$$
Then by a change of variables, for any bounded measurable set $B\in\mathbb R^n$ we have 
\begin{equation}
\label{measzero}
\int_{F^{-1}(B)} 1\cdot |DF(\vec t\,)|d\vec t = \int_B 1 \, d\vec v,
\end{equation}
so if $B$ has measure zero then the integral on the left side of \eqref{measzero} must vanish. But $|Df(\vec t\,)|$ is nonzero almost everywhere in $\vec t$ and so $F^{-1}(B)$ must have measure zero. 

In conclusion, \eqref{49.1} can hold for a set of $\vec t$ of positive measure only if $\vec \delta=0$, which also implies $\alpha=0$, so $\gamma_{l_0}=\gamma_{r_0}$ unless $\omega_1t\in \pi\mathbb Z$. Finally, suppose there is another $k_2\in\mathcal N_0$ such that $\omega_2\equiv \omega(k_2)\neq 0$. Then repeating the above argument gives that  \eqref{49.1} can hold for a set of $\vec t$ of positive measure only if $\vec \delta=0$, and $\gamma_{l_0}=\gamma_{r_0}$ unless $\omega_2t\in \pi\mathbb Z$. If $\omega_2/\omega_1\not\in\mathbb Q$ then the only time for which $\omega_1t, \omega_2t\in\pi\mathbb Z$ is $t=0$. So for $t\neq 0$ we have \eqref{eqlem2} for almost every $\vec t$.  

This completes the proof of Lemma \ref{lem:2}.\hfill $\qed$

\medskip
As the series in $n\ge 1$ in \eqref{term2eq2} converges uniformly in $\lambda$ we can interchange its summation and the limit $|\lambda|\rightarrow\infty$ and we can take the latter limit inside the multiple integral in \eqref{term2eq2} due to the Lebesgue dominated convergence theorem. Thus combining \eqref{bigWeyl}, \eqref{eqlem2} and \eqref{term2eq2} we arrive at
\begin{align}
\lim_{|\lambda|\rightarrow\infty} T_2  
= &  \sum_{l=1}^\nu \, \sum_{n\ge 1} i^n \int_{0\le t_n\le\cdots\le t_1\le t}  \omega_\s\big((P_l H_\s P_l)^n P_l A P_l\big)\omega_\r \big(W(e^{i\omega t}h )\big)\nonumber\\
= & \sum_{l=1}^\nu \omega_\s \big((e^{i t  P_l H_\s P_l}-\bbbone) P_l A P_l\big)\omega_\r \big(W(e^{i\omega t}h )\big),
 \label{T2}
\end{align}
where we resummed
$$
\sum_{n\ge 1} i^n \int_{0\le t_n\le\cdots\le t_1\le t}  (P_l H_\s P_l)^n = \sum_{n\ge 1} \frac{(i t)^n}{n!} (P_l H_\s P_l)^n = e^{i t  P_l H_\s P_l}-\bbbone.
$$

$\bullet$ We take the limit $|\lambda|\rightarrow\infty$ of $T_3$ (see \eqref{sum}). As
$$
\omega_\s\otimes\omega_\r\big( e^{it K} \big(B(\lambda)\otimes W(h)\big) e^{-itK}\mathcal D^* \big) = \omega_\s\otimes\omega_\r\Big( \big( \mathcal D e^{it K} \big(B(\lambda)^*\otimes W(-h)\big) e^{-itK}\big)^*\Big)
$$
we see that $T_3$ is the complex conjugate of $T_2$ with $B(\lambda)$ replaced by $B(\lambda)^*$ and $h$ replaced by $-h$. So we obtain from \eqref{T2}, 
\begin{align}
\lim_{|\lambda|\rightarrow\infty} T_3  
=  \sum_{l=1}^\nu \omega_\s \big( P_l A P_l (e^{-i t  P_l H_\s P_l}-\bbbone)\big)\omega_\r \big(W(e^{i\omega t}h ) \big).
 \label{T3}
\end{align}

\medskip

$\bullet$ We take the limit $|\lambda|\rightarrow\infty$ of $T_4$ (see \eqref{sum}). We have
\begin{align}
T_4 = & \omega_\s\otimes\omega_\r\big(\mathcal D e^{it K} \big(B(\lambda)\otimes W(h)\big) e^{-itK}\mathcal D^*\big)\nonumber\\
= &  \sum_{l_0,r_0=1}^\nu \, e^{-\frac i2 \lambda^2(\gamma^2_{l_0}-\gamma^2_{r_0}){\rm Im} \langle g, \frac{e^{-i\omega t}+it\omega}{\omega^2}g\rangle}   \nonumber\\
& \ \times \omega_\s\otimes\omega_\r\big(\mathcal D P_{l_0} AP_{r_0}  W\big((e^{it\omega}-1)(f_{l_0}-f_{r_0})+e^{i\omega t}h\big) \mathcal D^*\big).
\label{term4}
\end{align}
Next, from the definition of $\mathcal D$, \eqref{D}
\begin{align}
D P_{l_0}&AP_{r_0}  W\big((e^{it\omega}-1)(f_{l_0}-f_{r_0})+e^{i\omega t}h\big) \mathcal D^*\nonumber\\
=&  \sum_{n\ge 1}\sum_{m\ge 1} i^n (-i)^m \int_{0\le t_n\le\cdots\le t_1\le t}\int_{0\le s_m\le \cdots \le s_1\le t} \nonumber\\
& \times H_\s(t_n)\cdots H_\s(t_1) P_{l_0} AP_{r_0}  W\big((e^{it\omega}-1)(f_{l_0}-f_{r_0})+e^{i\omega t}h\big) H_\s(s_1)\cdots H_\s(s_m).
\label{term4eq2}
\end{align}
We use Lemma \ref{lem:1} to get
\begin{align}
&H_\s(t_n)\cdots H_\s(t_1) P_{l_0} AP_{r_0}  W\big((e^{it\omega}-1)(f_{l_0}-f_{r_0})+e^{i\omega t}h\big) H_\s(s_1)\cdots H_\s(s_m)\nonumber \\
&\, =  \sum_{l_1,\ldots,l_n=1}^\nu \sum_{r_1,\ldots,r_m=1}^\nu P_{l_n}H_\s P_{l_{n-1}} \cdots P_{l_1}H_\s P_{l_0} A P_{r_0} H_\s P_{r_1} \cdots P_{r_{m-1}}H_\s P_{r_m}\nonumber\\
& \quad \times  W\big((e^{it_n\omega}-1)(f_{l_n}-f_{l_{n-1}})\big) \cdots W\big((e^{it_1\omega}-1)(f_{l_1}-f_{l_0})\big)  W\big((e^{it\omega}-1)(f_{l_0}-f_{r_0})+e^{i\omega t}h\big)\nonumber\\
& \quad \times W\big((e^{is_1\omega}-1)(f_{r_0}-f_{r_1})\big) \cdots W\big((e^{is_m\omega}-1)(f_{r_{m-1}}-f_{r_m})\big)\nonumber\\
&\, =  \sum_{l_1,\ldots,l_n=1}^\nu \sum_{r_1,\ldots,r_m=1}^\nu e^{-i \Phi_4} P_{l_n}H_\s P_{l_{n-1}} \cdots P_{l_1}H_\s P_{l_0} A P_{r_0} H_\s P_{r_1} \cdots P_{r_{m-1}}H_\s P_{r_m}\nonumber\\
&\!\!\!\times  W\Big(\sum_{j=1}^n(e^{it_j\omega}-1)(f_{l_j}-f_{l_{j-1}})\big) +\sum_{k=1}^m (e^{is_k\omega}-1)(f_{r_{k-1}}-f_{r_k})+ (e^{i\omega t}-1)(f_{l_0}-f_{r_0})+e^{i\omega t}h \Big).
 \label{term4eq3}
\end{align}
The phase $\Phi_4$ comes from the commutation relations of the Weyl operators when we combine their products into a single one. It satisfies
\begin{equation*}
\Phi_4 = 0\qquad \mbox{if $l_n=\cdots =l_1=l_0=r_0=r_1=\cdots=r_m$.}
\end{equation*}
The expectation of the Weyl operator in \eqref{term4eq3} in $\omega_\r$ is,
\begin{align}
e^{-\frac14 \big\| \mathcal C^{1/2} \big[ \sum_{j=1}^n\frac{e^{it_j\omega}-1}{i\omega} \lambda (\gamma_{l_j}-\gamma_{l_{j-1}})g +\sum_{k=1}^m \frac{e^{is_k\omega}-1}{i\omega}\lambda (\gamma_{r_{k-1}}-\gamma_{r_k})g +\frac{e^{it\omega}-1}{i\omega}\lambda(\gamma_{l_0}-\gamma_{r_0})g +e^{i\omega t}h\big]\big\|^2_{L^2}}.
\label{normterm4}
\end{align}
By replicating the proof of Lemma \ref{lem:2} we see that unless $\gamma_{l_n}=\cdots=\gamma_{l_0}=\gamma_{r_0}=\cdots =\gamma_{r_m}$, we have that the limit of \eqref{normterm4} as $\lambda\rightarrow 0$ is zero, for all $t>0$ and all $t_1,\ldots,t_n,s_1,\ldots,s_m$ (except possibly a set of measure zero for the $(t_1,\ldots,s_m)$). Using this together with \eqref{term4}, \eqref{term4eq2} \eqref{term4eq3}, we arrive at
\begin{align}
\lim_{\lambda\rightarrow\infty} T_4 = & \sum_{n\ge 1}\sum_{m\ge 1} i^n (-i)^m \int_{0\le t_n\le\cdots\le t_1\le t}\int_{0\le s_m\le \cdots \le s_1\le t}\nonumber\\
& \sum_{l=1}^\nu \omega_\s\big( (P_l H_\s P_l)^n P_l A P_l (P_l H_\s P_l)^m\big) \omega_\r(W(e^{i\omega t}h)) \nonumber\\
= & \sum_{l=1}^\nu \omega_\s\big( (e^{it P_l H_\s P_l} -\bbbone) P_l A P_l (e^{-it P_l H_\s P_l}-\bbbone)\big)\ \omega_\r(W(e^{i\omega t}h)).
\label{T4}
\end{align}

Summing up the limits as $|\lambda|\rightarrow\infty$ of $T_1$ to $T_4$ according to \eqref{T1}, \eqref{T2}, \eqref{T3} and \eqref{T4}, we arrive at the result \eqref{thmeqn}. This completes the proof of Theorem \ref{thm2}.\hfill \qed

\subsection{Proof of Theorem \ref{thm2res}}
\label{sect:proof2res}

Theorem \ref{thm2res} is proved similarly to Theorem \ref{thm2}. We highlight the details that differ. Define a new operator $K$ (compare with \eqref{K})
\begin{equation*}
    K = H_{\r 1} +  H_{\r 2} + \lambda_2 G_2 \otimes \varphi_2(g_2),
\end{equation*}
so that
\begin{equation*}
    e^{itK}= e^{it H_{\r 1}} \sum_{l=1}^{\nu} P_l e^{it (H_{\r 2} + \lambda_2 \gamma_l \varphi_2(g_2))} .
\end{equation*}
Here, the $P_l$ and $\gamma_l$ are the spectral projections (of dimension $\ge 1$) and the distinct eigenvalues of $G_2$,
\begin{equation*}
G_2 = \sum_{l=1}^\nu \gamma_l P_l.
\end{equation*}
Formally the Dyson series reads
\begin{equation*}
    e^{i t H} e^{- i t K} = \bbbone + \mathcal{D}, \qquad \mathcal{D} = \sum_{n\ge 1} i^n \int_{0\le t_n\le\cdots\le t_1\le t} \widetilde{H}(t_n)\cdots \widetilde{H}(t_1)
\end{equation*}
where (as in Lemma \ref{lem:1})
\begin{align*}
    \widetilde{H}(t) &= e^{i t K} (H - K) e^{-i t K} \nonumber \\
    &=\sum_{l,r=1}^\nu P_l \left( H_\s + \lambda_1 G_1 \otimes \varphi_1(e^{i\omega t}g_1) \right)P_r  \nonumber \\
    &\qquad \times
    W_2 \big((e^{i\omega t}-1)(f_l-f_r)\big)  e^{-\frac i2 {\rm Im} \big\langle (e^{i\omega t} -1 -i\omega t)(f_l-f_r),(f_l+f_r)\big\rangle} .
\end{align*}
A technical difference with respect to the previous case (single reservoir) is that the presence of the field operator $\varphi_1$ makes $\widetilde{H}(t)$ an unbounded operator. The condition \eqref{r1regularA} makes sure the Dyson series converges in the weak sense, that is, when the state $\omega_{\r 1}$ is applied, as we explain below. As in the proof of Theorem \ref{thm2}, we set
\begin{eqnarray*}
\lefteqn{
\omega_\s\otimes\omega_{\r1}\otimes\omega_{\r2}\big(e^{it H} \big(A\otimes \bbbone\big) e^{-itH}\big) }\nonumber\\
& \qquad = & \omega_\s\otimes\omega_{\r1}\otimes\omega_{\r2}\big(e^{it K} \big(A\otimes \bbbone\big) e^{-itK}\big)\\
 && +\omega_\s\otimes\omega_{\r1}\otimes\omega_{\r2}\big(\mathcal D e^{it K} \big(A\otimes \bbbone\big) e^{-itK}\big)\\
& & +\omega_\s\otimes\omega_{\r1}\otimes\omega_{\r2}\big( e^{it K} \big(A\otimes \bbbone\big) e^{-itK}\mathcal D^* \big)\\
& & +\omega_\s\otimes\omega_{\r1}\otimes\omega_{\r2}\big(\mathcal D e^{it K} \big(A\otimes \bbbone\big) e^{-itK}\mathcal D^*\big)\\
&\qquad \equiv & T_1+T_2+T_3+T_4
\end{eqnarray*} 
and we take $|\lambda_2| \to \infty$ in each term. Here, $\bbbone$ is the identity operator acting on both reservoir Hilbert spaces. In the expression of $T_1$ the free Hamiltonian of the first reservoir $H_{\r1}$ drops out (the propagator commutes with $A\otimes\bbbone$) and the expression reads exactly as \eqref{term1} with $h=0$ and $\lambda_2$ in place of $\lambda$. Therefore, thanks to Lemma \ref{lem:1}, we obtain (see also \eqref{T1} with $h=0$),
\begin{equation}
\label{2resT1}
    \lim_{|\lambda_2| \to \infty} T_1 = \sum_{l=1}^\nu \omega_\s(P_l A P_l).
\end{equation}
Next we analyze $T_2$. Proceeding as in the derivation of \eqref{term2eq2} we now obtain
\begin{align}
    T_2 &= \omega_\s\otimes\omega_{\r1}\otimes\omega_{\r 2} \Bigg( \sum_{n\ge 1} \,\sum_{l_0,\ldots,l_n, r_0=1}^\nu \!\!\!  i^n \int_{0\le t_n\le\cdots\le t_1\le t} \!\!\!\!\! e^{-\frac i2 \Phi(\lambda_2^2)} P_{l_n}\widehat{H}(t_n) P_{l_{n-1}}\cdots P_{l_1}\widehat{H}(t_1) P_{l_0} A P_{r_0}\nonumber\\
 &\quad  \times W_2 \big( \sum_{j=1}^n (e^{it_j\omega}-1)(f_{l_j}-f_{l_{j-1}}) + (e^{it\omega}-1)(f_{l_0}-f_{r_0}) \Big) \Bigg) ,
 \label{dysonT2}
\end{align}
where 
$$
\widehat{H}(t)= H_\s + \lambda_1 G_1 \otimes \varphi_1(e^{i\omega t}g_1).
$$
We now show that the right side of \eqref{dysonT2} is well defined. We have $|\omega_{\r 2}(W_2(f))|\le 1$ for any $f\in L^2({\mathbb R^3},d^3k)$ and 
\begin{equation*}
    \left| \omega_\s\otimes\omega_{\r1} \left( P_{l_n}\widehat{H}(t_n) P_{l_{n-1}}\cdots P_{l_1}\widehat{H}(t_1) P_{l_0} A P_{r_0} \right) \right| \leq \|A\| 2^n C^n B_n,
\end{equation*}
where $C=\max\{\| H_\s \|,|\lambda_1|\, \|G_1 \|\}$ and $B_n$ is as in \eqref{r1regularA}. Then, due to \eqref{r1regularB},
\begin{align}
&\sum_{n\ge 1} \,\sum_{l_0,\ldots,l_n, r_0=1}^\nu 
  \int_{0\le t_n\le\cdots\le t_1\le t} \left| \omega_\s\otimes\omega_{\r1} \left( P_{l_n}\widehat{H}(t_n) P_{l_{n-1}}\cdots P_{l_1}\widehat{H}(t_1)  P_{l_0} A P_{r_0} \right)  \omega_{\r2} \Big(W_2 \big( \lambda_2 h_n \big) \Big)  \right|\nonumber\\
&\qquad\quad \le \|A\| \sum_{n\ge 1} \nu^{n+2}  \frac{(2Ct)^n}{n!}B_n <\infty.
\label{series}
\end{align}
Furthermore, the series $\sum_{n\ge 1}$ on the left side in \eqref{series} converges uniformly in $\lambda_2$ and so we can pull the limit $|\lambda_2|\rightarrow\infty$ inside the sum in \eqref{dysonT2} to obtain,
\begin{align}
    \lim_{|\lambda_2| \to \infty} T_2 &= \omega_\s\otimes\omega_{\r1} \Big( \sum_{l=1}^\nu \sum_{n\ge 1} i^n \int_{0\le t_n\le\cdots\le t_1\le t}  P_{l}\widehat{H}(t_n) P_{l}\cdots P_{l}\widehat{H}(t_1) P_{l} A P_{l} \Big) \nonumber \\
    &= \omega_\s\otimes\omega_{\r1} \Big( \sum_{l=1}^\nu P_l \big(e^{it (  H_{\r1} + P_l H_\s P_l + \lambda_1 P_l G_1 P_l \otimes \varphi_1(g_1))}e^{-it H_{\r1}} -\bbbone \big) P_{l} A P_{l} \Big).
\label{2resT2}
\end{align}
The first equality in \eqref{2resT2} is obtained as before using Lemma \ref{lem:2} which constrains $r_0=l_0=l_1=\ldots=l_n$ (then also making the phase $\Phi$ in \eqref{dysonT2} vanish). To arrive at the second equality in \eqref{2resT2} we used the Dyson series expansion
\begin{equation*}
    e^{it (  H_{\r1} + P_l H_\s P_l + \lambda_1 P_l G_1 P_l \otimes \varphi_1(g_1))}e^{-it H_{\r1}} = \bbbone + \sum_{n\ge 1} i^n \int_{0\le t_n\le\cdots\le t_1\le t}  P_{l}\widehat{H}(t_n) P_{l}\cdots P_{l}\widehat{H}(t_1)P_l .
\end{equation*}
Just as in the single reservoir case, the term $T_3$ can be obtained from $T_2$ by a suitable complex conjugation (see before \eqref{T3}),
\begin{equation}
\label{2resT3}
    \lim_{|\lambda_2| \to \infty} T_3 = \omega_\s\otimes\omega_{\r1} \Big( \sum_{l=1}^\nu P_{l} A P_{l} (e^{it H_{\r1}} e^{-it (  H_{\r1} + P_l H_\s P_l + \lambda_1 P_l G_1 P_l \otimes \varphi_1(g_1))} -\bbbone ) P_l \Big) .
\end{equation}

Finally, the term $T_4$ is again treated similarly. We need to make sure the double Dyson series in $n$ and $m$ converges weakly (compare with \eqref{term4eq2}). It suffices check that the absolute double series converges, which is shown by noticing that for any $\alpha \in \mathbb{R}$,
\begin{equation*}
\sum_{n \geq 0} \sum_{m \geq 0} \frac{\alpha^{n+m}}{n! \, m!} B_{n+m}= \sum_{n \geq 0} \sum_{k \geq n} \frac{\alpha^{k}}{n! (k-n)!} B_{k} = \sum_{k \geq 0} \frac{\alpha^{k}}{k!} B_{k} \sum_{n = 0}^k \binom{k}{n} = \sum_{k \geq 0} \frac{2^k \alpha^{k}}{k!} B_{k} < \infty.
\end{equation*}
We then obtain,
\begin{align}
    \lim_{|\lambda_2| \to \infty} T_4 &= \omega_\s\otimes\omega_{\r1} \Big( \sum_{l=1}^\nu P_l (e^{it (  H_{\r1} + P_l H_\s P_l + \lambda_1 P_l G_1 P_l \otimes \varphi_1(g_1))}e^{-it H_{\r1}} -\bbbone ) P_{l} A P_{l} \nonumber \\
    &\quad \times (e^{it H_{\r1}} e^{-it (  H_{\r1} + P_l H_\s P_l + \lambda_1 P_l G_1 P_l \otimes \varphi_1(g_1))} -\bbbone ) P_l \Big) .
    \label{2resT4}
\end{align}
Finally, summing the four contributions \eqref{2resT1}, \eqref{2resT2}, \eqref{2resT3} and \eqref{2resT4} gives the result \eqref{lim2res}. This completes the proof of Theorem \ref{thm2res}.\hfill $\qed$

\bigskip

{\bf Acknowledgements.} The authors are grateful to two anonymous referees. Their insights and comments are much appreciated. MM acknowledges the support of a Discovery Grant from the Natural Sciences and Engineering Council of Canada (NSERC). The work of SM is funded under the Horizon Europe research and innovation program through the MSCA project ConNEqtions, n.~101056638. SM received funding from the GNFM of INdAM to participate in the program Indam Quantum Meetings 2022 (IQM22) in Milan, where the authors could meet and discuss about topics related to this article. SM is grateful to the Department of Mathematics and Statistics of Memorial University of Newfoundland for the hospitality while working on this project.

\end{document}